\documentclass[11pt]{article}
\usepackage[margin=1in]{geometry}
\usepackage[T1]{fontenc}
\usepackage{graphicx}
\usepackage{tabularx}
\newcolumntype{Y}{>{\raggedright\arraybackslash}X}
\usepackage{amsmath,amssymb}
\usepackage{booktabs}
\usepackage{array}
\usepackage[version=4]{mhchem}
\usepackage{siunitx}
\usepackage{xcolor}
\usepackage[numbers,sort&compress]{natbib}
\usepackage{hyperref}

\graphicspath{%
  {Figures/}{Figures/figs/}{Figures/paper_figures/}%
  {Figures/figures/Li_topology/}{Figures/figures/Na_topology/}%
  {Figures/figures/bottleneck_topology/}{Figures/figures/bottleneck_topology/Na/}%
  {Figures/figures/ph_gles/}{Figures/figures/redundancy_degeneracy/}%
  {Figures/figures/vacancy_percolation/}{Figures/figures/vacancy_percolation/Na/}}

\title{\vspace{-2em}Topo-Spectral Percolation Descriptors for Mechanistic\\
Ion Transport Pathways from Static Crystal Structures}

\author{Diptendu Roy, Chiku Parida, Juan Mar\'ia Garc\'ia-Lastra, Arghya Bhowmik\\[2pt]
\small Technical University of Denmark, Department of Energy Conversion and Storage, Lyngby, 2800, Denmark\\
\small Correspondence: dipro@dtu.dk, arbh@dtu.dk}
\date{}

\begin{document}
\maketitle

\section*{Significance Statement}
Studies of ion transport usually aim to predict the transport rate and the migration mechanism from ab initio or force-field molecular dynamics trajectories computed over a range of temperatures. Ion transport has two separate components. The static framework determines the transport mechanism, which includes the diffusion routes, the bottlenecks, the transport dimensionality, and the vacancy fraction needed for mobility. Dynamical factors determine the kinetics along this structurally defined pathway. We show that the diffusion mechanism can be obtained from a single static structure in seconds. Our method computes physics-based migration barriers on a periodic network and analyzes the network through its topology and its graph spectrum. We validate it against ab initio molecular dynamics and neutron diffraction for eight ion-conducting materials. These results will enable high-throughput analysis of ion-conduction mechanisms in materials libraries.

\begin{abstract}
\noindent
Ion transport controls the macroscopic performance of solid electrolytes, battery electrodes, ion-selective membranes, electrolysis ceramics, mixed conductors, porous sorbents, and biological channels. In each case, ion transport is determined by structural characteristics, namely the available diffusion pathways and the specific bottleneck atoms that limit the ionic jumps. These are intrinsic properties of the crystalline framework, whereas temperature, carrier concentration, and correlated motion govern the transport rate. Ab initio molecular dynamics (AIMD) resolves the mechanism directly but is too costly for materials screening, and most surrogate models give only a single transport number. Here we present a calibration-free method for determining the mechanism, based on Topo-Spectral Percolation Descriptors (TSPD). TSPD obtains the transport mechanism from a single static structure in seconds. It constructs a periodic migration network whose edge barriers are computed from physics-based energetics. We then analyze the barrier-threshold topology of this network and the spectrum of its barrier-weighted graph Laplacian. To establish its validity for computational solid-state ionics, we test TSPD against AIMD ion densities and neutron diffraction data for eight materials, including cathodes and solid electrolytes, whose transport ranges from one- to three-dimensional. TSPD reproduces the established mechanism in every case and correctly identifies structures in which geometric openness does not imply viable ion percolation. The method is most informative for pathway-limited, direction-dependent transport, and it complements full dynamical simulations for strongly cooperative or nearly isotropic transport.
\end{abstract}

\section{Introduction}
The diffusion routes available to a mobile ion in a crystal are determined mainly by the equilibrium framework, and molecular dynamics is used to sample these routes. The structural features of a diffusion route give a mechanism-level description of transport. These features are its connectivity, its transport dimensionality, the rate-limiting bottleneck with its neighboring framework atoms, and the vacancy fraction needed for long-range mobility. They relate directly to materials design strategies, such as the choice of substitution sites, carrier-introducing dopants, and phase selection. The same features apply to a wide range of ion-transport materials, from solid electrolytes and electrodes to membranes, electrolysis ceramics, mixed conductors, porous sorbents, and ion-selective solids.

The crystal framework determines the \emph{transport mechanism}, that is, the percolating routes and bottlenecks, can be extracted from the static structure. The dynamics then determine the \emph{transport rate}; the conductivity or diffusivity depends on temperature, carrier concentration, and correlated motion. The mechanism is, however, already contained in a single static structure. Our central claim is that it can be recovered without expensive AIMD or force-field MD, and without any empirical fitting or parameters calibrated on transport observables.

AIMD is the reference method for this problem. It resolves the density of the migrating ion directly and establishes the mechanism reliably.\citep{Mo2012,He2017,Bachman2016} Its high computational cost, however, limits its use for high-throughput screening. A converged trajectory covers one composition at one temperature and is expensive, so the approach does not scale to materials databases. Faster static methods have therefore been developed, including frozen-lattice pinball models,\citep{Kahle2018,Kahle2020} geometric void analysis,\citep{He2020} bond-valence site-energy mapping,\citep{Adams2006,Chen2019} charge-density pathfinding,\citep{Rong2015} and electrostatic-corrugation descriptors with Born--Mayer repulsion.\citep{FranceLanord2019,Gustafsson2024,Mace2019} Topological and graph-based methods have also been used. Persistent homology of AIMD trajectories identifies the loops associated with concerted hops, but it requires the trajectory.\citep{Sato2023} The topology of the procrystal electron density locates channels that guide a subsequent first-principles study, but it does not report a mechanism.\citep{Golov2022} The closest earlier approach is a migration graph whose edges are weighted by individually computed hop barriers, with percolation found as the lowest-cost cycle. However, this approach requires an expensive barrier calculation for each hop.\citep{Shen2023} When topology is used as a machine-learning feature, or Betti curves of the electron density are computed, the result is a similarity score or a property and not a route.\citep{Chen2025,Szymanski2025}

Most static surrogates reduce a structure to a single calibrated transport number. This hides the separation between the mechanism, which is determined by the structure, and the rate, which is driven by the dynamics. An open network conducts poorly when carriers are absent, whereas a narrow bottleneck conducts well once relaxation and correlated motion dominates. We therefore target the transport mechanism directly. The mechanism contains the information relevant for design even when the absolute rate can still be adjusted.

We introduce Topo-Spectral Percolation Descriptors (TSPD), which recover the migration mechanism from static structures. TSPD constructs a periodic migration network for the mobile ions. Each edge carries a migration-barrier profile computed from parameter-free physics. The profile contains an overlap-penalized short-range repulsion constrained by mechanical equilibrium, screened electrostatics with a dielectric constant obtained from the framework polarizability, an induced-dipole response, and a chemical-softness term from HSAB theory that distinguishes hard oxides from softer sulfides and halides. We analyze the same network in two ways. A barrier-threshold topological analysis gives the transport dimensionality, the percolating route as a minimax path, the rate-limiting bottleneck with its neighboring framework atoms, and the redundancy of this route, measured by the barrier increment to the next available alternative. A barrier-weighted spectral analysis also measures the redundancy of the route. Both analyses order the edges by migration barrier. This barrier-based filtration therefore defines the transport mechanism through an energetic topology and not through static spatial descriptors, which distinguishes the method from earlier treatments based on geometric topology and proximity graphs.

Because TSPD gives the transport mechanism, it can be validated at the mechanism level by comparing the predicted routes and bottlenecks with independent ion-density data. As a proof of concept, we have studied eight materials that cover cathode hosts and solid electrolytes: \ce{LiFePO4}, \ce{LiCoO2}, \ce{Li10GeP2S12}, cubic and tetragonal \ce{Na3PS4}, \ce{Li6PS5Cl}, and two ordered oxides in which transport is dominated by cooperative or defect-mediated processes, tetragonal \ce{Li7La3Zr2O12} and $\gamma$-\ce{LiAlO2}. We compare the TSPD results with published AIMD ion-density maps and with site populations derived from neutron diffraction. For all structures in this set, TSPD reproduces the established transport mechanisms and correctly identifies the structures that appear geometrically open but have no viable transport pathway. Each analysis takes seconds per structure on a single processor.
The main contribution of this work is to show that the transport mechanism can be treated as a separable, structurally defined quantity that can be obtained without expensive computation or empirical parameterization. This provides a basis for future inverse design and for analyses of order--disorder effects. The rest of the paper describes the network and its energetics, defines the topological descriptors, and evaluates eight representative structures along the ladder of transport dimensionality.

\begin{figure*}[t]
\centering
\includegraphics[width=\textwidth]{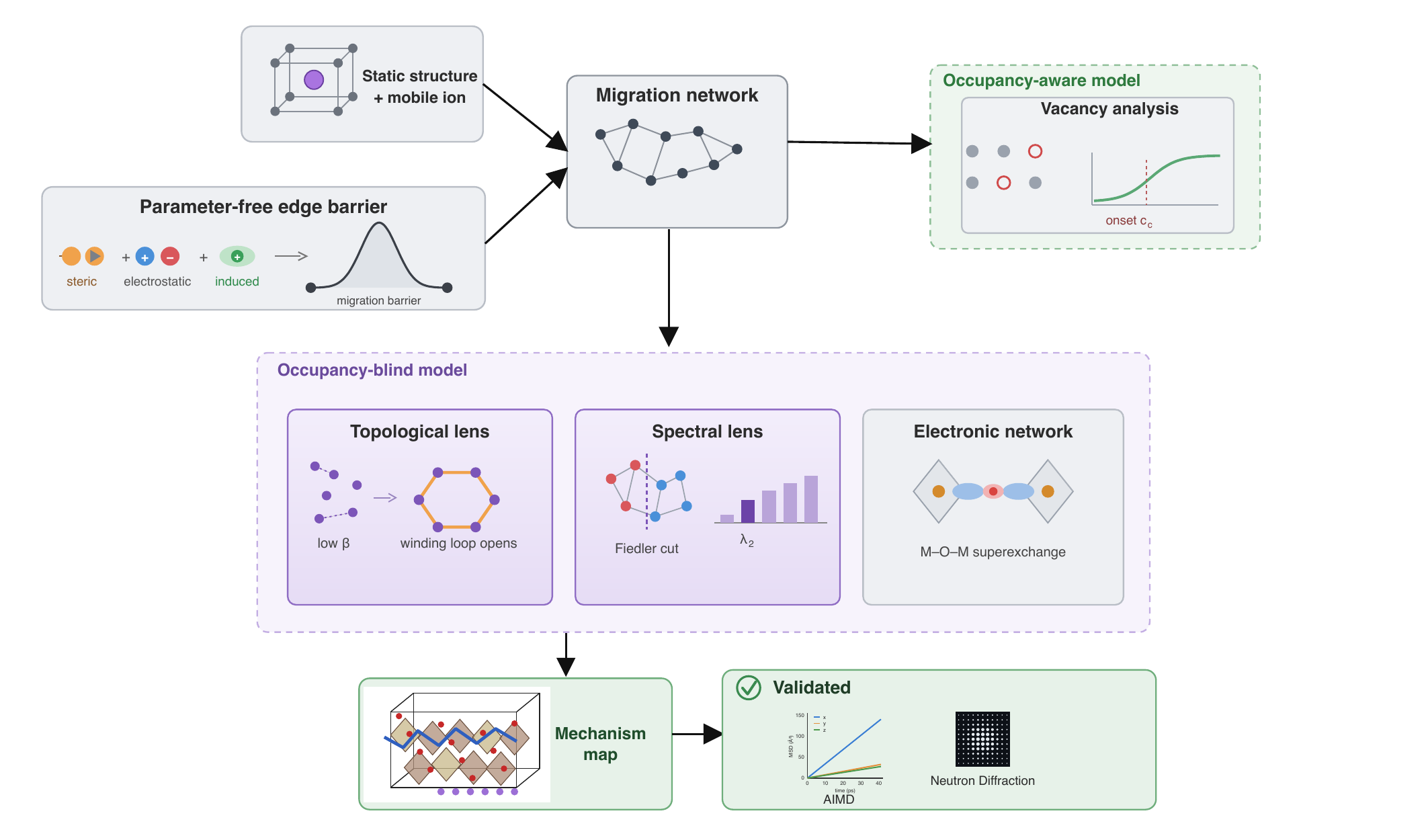}
\caption{The TSPD workflow} 
\label{fig:workflow}
\end{figure*}

\section{Theory and Method}
Here we summarize the framework and the main quantities. The mathematical definitions of all terms and descriptors are given in the Supporting Information (SI).

\subsection{The migration network}
Starting from a crystal structure with its site occupancies and a specified mobile species, we construct a
periodic migration graph (SI Section~S1). The nodes are the candidate
mobile-ion sites and the interstitial voids that connect them, and the edges are
the geometrically allowed hops between neighboring nodes. A filter that takes species and bonding into account keeps framework atoms out of the graph. A cation bonded within an anion polyhedron is
treated as part of the framework regardless of its site occupancy, so that \ce{Ge} and \ce{P} in a thiophosphate, or \ce{Fe} and \ce{P} in an olivine, never enter the mobile network. We retain two networks. A coarse skeleton containing only the resting sites of the mobile species is used for the connectivity and transport dimensionality, which are crystallographic and
independent of the choice of origin. The void-augmented network describes the full curved route that an ion follows
between these sites. The interstitial voids are generated by a Voronoi construction, which gives the same set of voids when the cell origin is shifted. A void is added to a hop only if the direct neck between the resting sites is already passable, so the voids annotate open
routes but do not create new ones.

\subsection{Parameter-free edge energetics}
Each edge carries a migration-barrier profile made up of three physical contributions, all evaluated without empirical fitting (SI Section S2). At first, we have considered an overlap-penalized Born–Mayer steric term, which sums the repulsion between the mobile ion and each framework neighbor using the explicit ion–framework distances. Its prefactor is fixed by requiring force balance with the electrostatic term at the equilibrium coordination. Second term is a screened-Ewald term for the electrostatic interaction of the mobile ion with the framework charges, with a static dielectric constant estimated from the framework polarizabilities through a Clausius–Mossotti relation. The third is an induced-dipole term, which describes the polarization response of the framework anions. In addition, anion softness is included through hard–soft acid–base (HSAB) chemistry by scaling the Born–Mayer range parameter. As a result, soft sulfide and chloride frameworks have a softer, longer-ranged repulsive barrier than hard oxide frameworks.

We sample the profile along an edge $\{u,v\}$ at the positions $P(t)=(1-t)\,c_u+t\,c_v$. The edge barrier $B_{uv}$ is the maximum energy along this path, measured from the global minimum energy of the mobile sites,
$E_0$,
\begin{equation}
B_{uv}=\max_{t\in[0,1]}\Big[E_{\mathrm{ster}}(P(t))+E_{\mathrm{elec}}(P(t))
+E_{\mathrm{ind}}(P(t))\Big]-E_0,
\label{eq:edgebarrier}
\end{equation}
where $E_{\mathrm{ster}}$, $E_{\mathrm{elec}}$, and $E_{\mathrm{ind}}$ are the steric,
screened-electrostatic, and induced-dipole terms parameterized in SI Section~S2. Referencing all energies to $E_0$ removes the systematic offset due to coordination. Because every energy is measured from the same baseline, spurious variations in local site stability caused by different coordination numbers cancel, and the barriers isolate the actual kinetic obstacles. At the point of maximum energy, i.e., the saddle, we split the energy into its steric and electrostatic parts, which defines the geometric and
electrostatic objectives used in the decomposition. The barrier is ordinal. It ranks
crystallographic directions and locates the bottleneck within a structure, but it is not a real
transport rate.

\subsection{Three objectives on one network}
We analyze the same barrier-weighted network with three objectives. The \emph{unified}
objective uses the full energetics and defines the route that we report as the prediction. The
\emph{geometric} objective keeps only the steric clearance, and the \emph{electrostatic}
objective keeps only the screened electrostatics. Comparing the three routes for a given material shows which of these factors contributes most to its bottleneck (SI Section~S2).

\subsection{Descriptors}
From the two analyses we compute a compact set of descriptors on the periodic network of mobile ions (SI Sections~S3--S8). Both analyses order the edges by migration barrier and not by geometry. The
topological analysis is based on a barrier filtration. Let
$\hat G_\beta=\{\,\{u,v\}:B_{uv}\le\beta\,\}$, with $B_{uv}$ from Eq.~\eqref{eq:edgebarrier}, be the subgraph of edges that are passable at threshold
$\beta$. The winding birth along lattice direction $a$ is the threshold at which $\hat G_\beta$
first contains a cycle that winds along $a$,
\begin{equation}
\Delta_a=\min\{\beta:\hat G_\beta\ \text{admits a periodic cycle winding along } a\}.
\label{eq:winding}
\end{equation}
The transport dimensionality is the number of axes with finite $\Delta_a$ (Eq.~\eqref{eq:winding}), and the ordinal
anisotropy is $\max_a\Delta_a-\min_a\Delta_a$. Geometrically, the winding birth is realized by the minimax (bottleneck) path, and the rate-limiting edge of this path defines the barrier for each axis,
\begin{equation}
B_{\mathrm{perc}}^{(a)}=\min_{\pi\in\Pi_a}\ \max_{\{u,v\}\in\pi}B_{uv}=\Delta_a,
\label{eq:minimax}
\end{equation}
so the topological analysis and the route coincide by construction. The barrier of the material is $B_{\mathrm{perc}}=\min_a B_{\mathrm{perc}}^{(a)}$. For the bottleneck we report its saddle coordinate, its symmetry multiplicity, and the framework atoms that define it (species and distances). Redundancy is the availability of alternative routes within a small barrier increment above the bottleneck. We report two readings of the same curve. The redundancy spectrum
$N(\Delta B)$ counts the windows with barriers within $\Delta B$ of
$B_{\mathrm{perc}}$; it fixes the energy and counts routes. The scalar $\Delta B_1$ fixes
the route and gives the energy. We remove the orbit of the rate-limiting hop from the network, solve
the minimax percolation of Eq.~\eqref{eq:minimax} again, and record the barrier increment
of the next percolating route (SI Section~S4). A small $\Delta B_1$ indicates a network that can reroute at low cost when it is obstructed, and a large value indicates a route with no
affordable alternative.

The spectral analysis uses the graph Laplacian. With nonnegative edge weights $w_{uv}$
that decrease with $B_{uv}$ and the weighted degree $D_{uu}=\sum_v w_{uv}$, the Laplacian
\begin{equation}
L=D-W,\qquad W_{uv}=w_{uv},
\label{eq:laplacian}
\end{equation}
is positive semidefinite, with eigenvalues $0=\lambda_1\le\lambda_2\le\cdots$. The algebraic
connectivity $\lambda_2$ of the resting-site network measures how densely the mobile sublattice is cross-linked, and through the Cheeger inequality it gives a lower bound on the minimum cut (Eq.~\eqref{eq:laplacian}). A saddle-cut test checks whether the minimax saddle lies on the spectral minimum cut
(SI Section~S5). Redundancy, as defined above, is the availability of an alternative route within a small
barrier increment of the bottleneck. The orbit of the rate-limiting hop is removed from the
network, the minimax percolation of Eq.~\eqref{eq:minimax} is solved again, and the
increment $\Delta B_1$ to the next percolating route is recorded (SI Section~S4).
The degeneracy is the symmetry multiplicity of the rate-limiting saddle, i.e., the size of its
orbit under the space-group operations of the crystal (SI Section~S4). These two quantities are independent. Degeneracy counts identical copies of one route, whereas redundancy gives the cost of a different route. A first-homology analysis on the same barrier filtration distinguishes
transport cycles, which close through a periodic image, from contractible cycles, which
close within a cell. By construction, the first transport cycle is born at the minimax
percolation threshold of Eq.~\eqref{eq:minimax}, because both orderings
are induced by the same edge barrier. The homology therefore provides an internal consistency
check on the bottleneck and the transport dimensionality and not an independent
measure.
An electronic network, weighted by the Goodenough--Kanamori superexchange geometry and analyzed with a Bloch spectral descriptor, is used to classify the material as an ideal electrolyte, a mixed conductor, or an insulator (SI Section~S7).

All descriptors introduced so far are computed occupancy blind. Every crystallographic site of the mobile species
is treated as an available node, so the descriptors give the geometric connectivity provided by the framework,
independent of how many mobile-ion sites are occupied. The vacancy response is the occupancy-aware
complement. In a global analysis, mobile ions are removed at random to determine the critical vacancy fraction
$c_c$ at which each axis  first percolates. In a per-class analysis, one symmetry-inequivalent class of mobile sites is vacated at a time while the others remain fully occupied. This shows whether a specific class alone enables percolation, gives its threshold fraction, and ranks the classes by their marginal effect. To make the result independent of the origin, the site classes are grouped with a frame-invariant fingerprint of the local environment and not with spglib orbits. A framework that is three-dimensionally connected in the occupancy-blind analysis can therefore remain non-conducting at full occupancy and start to conduct only when vacancies are introduced (SI Section~S8, Table~S3). 

\subsection{Route serialization as a validation gate}
The predicted route is stored as an ordered list of nodes with the periodic images
unwrapped, together with the coordinate frame and a crystallographic label for each
node (SI Section~S9). This information is needed for an unambiguous check. Aggregate descriptors can look reasonable even when a route is assigned incorrectly, and fractional
coordinates compared between different origin settings can appear to pass through framework atoms when they do not. Before any bottleneck or anisotropy result is reported, we verify in the correct frame that each node belongs to the mobile
sublattice or to a genuine interstitial site.

\subsection{Validation protocol}
For each structure we compute the route and compare it with the mechanism reported from
published AIMD, NEB, and neutron diffraction data. The comparison covers the path geometry, the bottleneck position, the framework
species that form the bottleneck, the transport dimensionality, and the vacancy requirement. We do not compare any TSPD barrier or
score with an experimental diffusivity or transport rate. The ordinal barrier
is used only to order directions and for anchoring within a structural family.
The structures, datasets, and computational environment are listed in SI Section~S10, and the validation status of each structure is given in SI Table~S4.

\section{Results}
We use the eight structures as a benchmark set to test the transport-mechanism concept along the ladder of transport dimensionality, from the one-dimensional olivine to the
three-dimensional electrolytes and the ordered oxides that conduct only by cooperative or defect-mediated migration. Two structures are analyzed in detail as representative cases, the redundant electrolyte \ce{Li10GeP2S12} and the one-dimensional cathode \ce{LiFePO4}, because they serve as reference cases for the bottleneck, redundancy, and vacancy analyses. The other six structures are introduced in the following sections where they illustrate specific mechanistic variations, and full details for each structure are given in the SI (SI Section~S13 and Tables~S1--S3). Two quantities are used throughout: $B^\ast$ and the percolation saddle. The direct-hop barrier $B^\ast$ (SI Eq.~S7) is the lowest barrier of a single hop between two mobile sites on the resting-site network, evaluated without void decoration. The percolation saddle is the highest point of the void-refined route along an open axis. The ordering of the two is diagnostic. If $B^\ast$ lies below the saddle, the direct hop along the channel is the low-barrier step. If $B^\ast$ lies well above the saddle, the route passes through an interstitial intermediate.

This criterion is a rigorous consequence of the definitions and does not rely on a correlation. A route made only of
direct hops between mobile-ion sites has a barrier of at least $B^\ast$, because $B^\ast$ is
the minimum over all such hops and a minimax route is limited by its highest-barrier
step. If $B^\ast$ exceeds $B_{\mathrm{perc}}$, the percolating route is
lower in energy than any direct hop available to it. This is possible only if
the route does not consist of direct hops, so it must pass through a site that is not a
resting site of the mobile ion. The inequality thus proves that an intermediate position exists, one that is not a
crystallographic resting site of the mobile species. It does so without reference to the geometry
of the route and from the static structure alone. In the opposite case, $B^\ast$ below $B_{\mathrm{perc}}$, transport runs along the direct channel and the network bottleneck lies elsewhere. This argument works only because $B^\ast$ is a minimum; the maximum has no such meaning. Since $B_{uv}$ is evaluated along the straight line between the two sites, a longer edge crosses more of the framework, and the largest
direct-hop barriers belong to lines that pass through a framework anion. In \ce{LiCoO2} the edge with the largest barrier is $5.66$~\AA{} long, has a clearance of $-1.41$~\AA{} at its narrowest point. The maximum is therefore set by the neighbor cutoff, which admits the longest edges, and not by any hop that the ion actually attempts. It exceeds $B_{\mathrm{perc}}$ for every structure in the set.

Two barriers are used throughout. $B_{\mathrm{perc}}$ is the highest barrier on the
lowest-barrier spanning route of the void-augmented network (Eq.~\eqref{eq:minimax}),
and $B^\ast$ (SI Eq.~S7) is the lowest barrier among direct hops between resting sites
of the mobile species. A route made only of direct hops cannot lie below $B^\ast$, so
$B^\ast>B_{\mathrm{perc}}$ proves that the percolating route passes through a position
that is not a resting site. The converse is not conclusive: $B^\ast<B_{\mathrm{perc}}$
shows only that the bottleneck is not the lowest direct hop, and the route may still
cross interstitial positions.

\subsection{Parameter-free energetics select the route}
\begin{figure}[htbp]
    \centering
    \includegraphics[width=\textwidth]{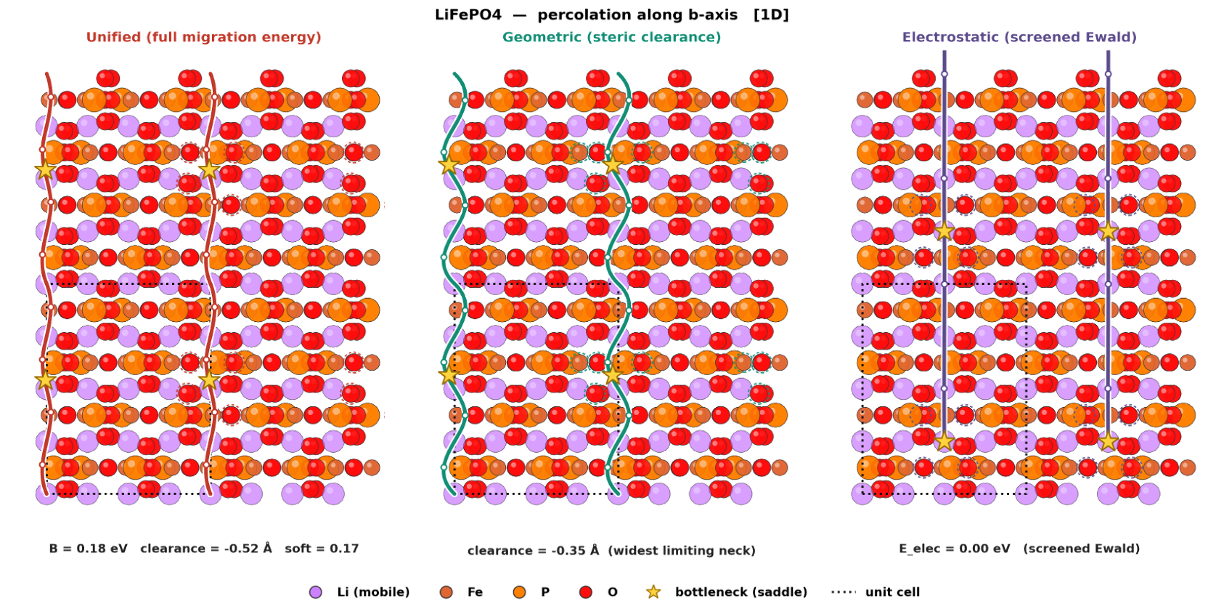}\\[8pt]
    \includegraphics[width=\textwidth]{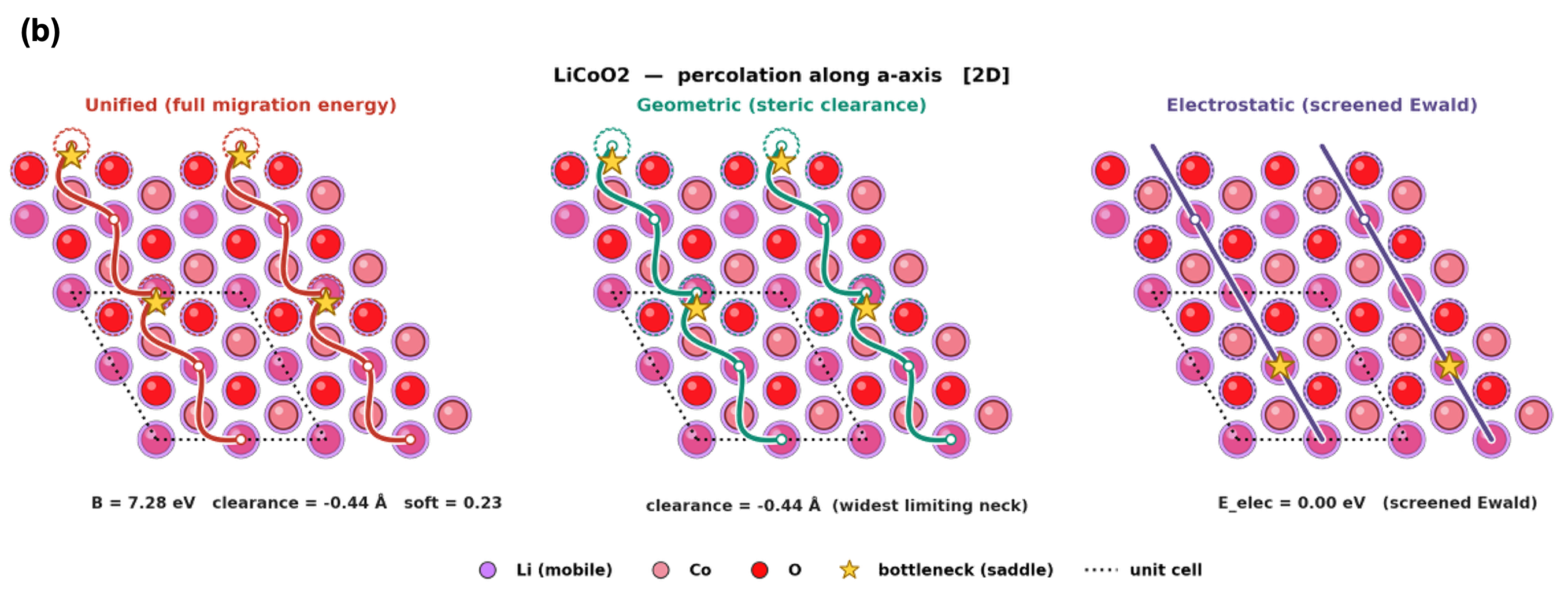}\\[8pt]
    \includegraphics[width=\textwidth]{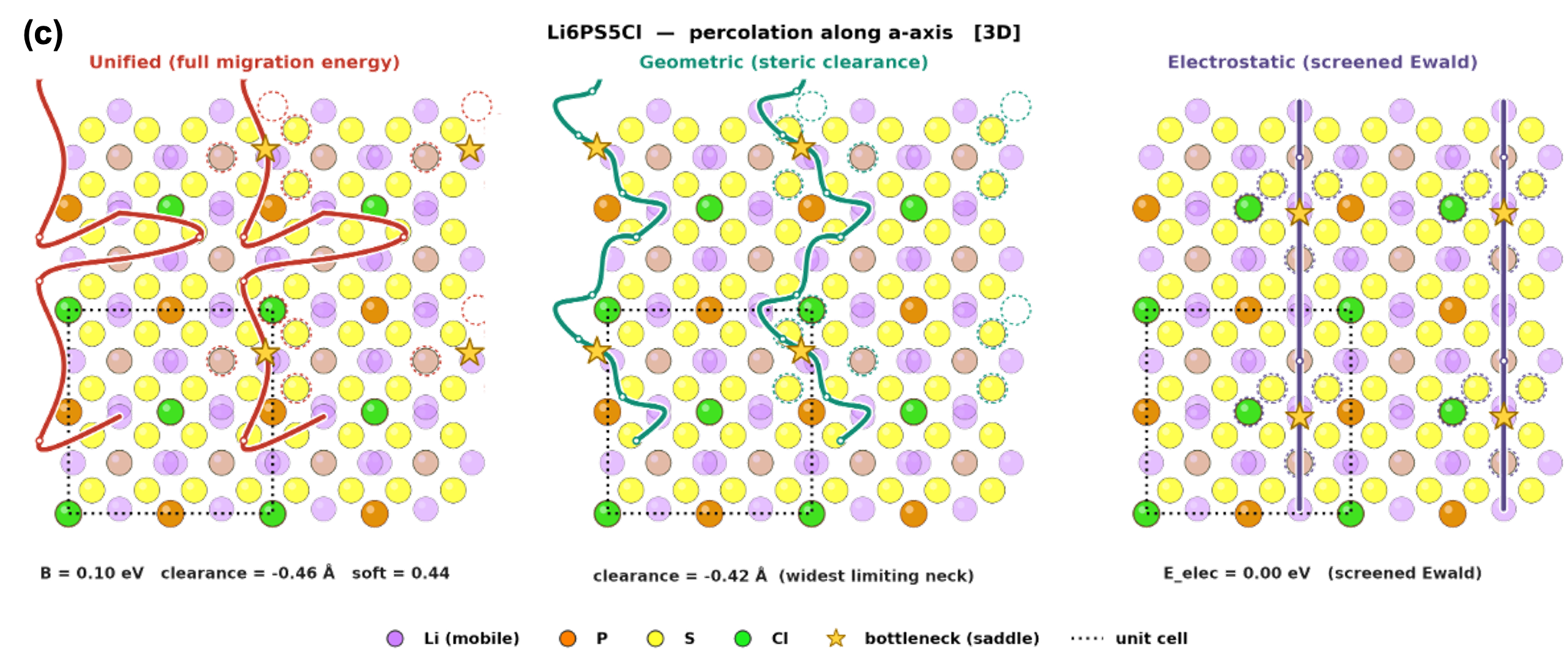}
    \caption{Decomposition of the parameter-free energetics along the ionic migration pathways: (a) \ce{LiFePO4}, (b) \ce{LiCoO2}, and (c) \ce{Li6PS5Cl}.}
    \label{fig:decomp_main}
\end{figure}

We find that parameter-free energetics, without any empirical fitting, are sufficient to select the correct migration route. Figure~\ref{fig:decomp_main} shows the decomposition of the route into the unified, geometric, and electrostatic objectives for three structures.

In \ce{LiFePO4} the steric, electrostatic, and induced-dipole contributions at the $b$-axis saddle are $27.0$, $-1.4$, and $-4.6$~eV, respectively, before subtraction of the site baseline $E_0$ of Eq.~\eqref{eq:edgebarrier}. The route is therefore set by the steric Born-Mayer wall, and electrostatics and polarization lower its height. The geometric route coincides with the
unified route, whereas the electrostatic-only route deviates from it, and Figure~\ref{fig:decomp_main} confirms that the path is selected by geometry and not by electrostatics. \ce{LiCoO2} behaves in the same way. Its saddle decomposes into $25.6$, $-0.16$, and $-5.9$~eV, and the unified route curves through
the interstitial while the electrostatic-only route is straight. \ce{Li6PS5Cl} represents the soft-anion limit. Its induced-dipole softening parameter, $0.55$, is the largest in the set, so the longer-ranged sulfide wall gives the least sterically hindered route among the structures considered (SI Figures~S1 and~S2). For the high-symmetry channel of cubic \ce{Na3PS4}, the steric, electrostatic, and induced-dipole contributions at the saddle are $10.8$, $-0.12$, and $-4.1$~eV. Here the geometric and electrostatic routes coincide instead of diverging, so the route is unambiguous. The opposite limit is $\gamma$-\ce{LiAlO2}, where the electrostatic term at the saddle is repulsive ($+2.3$ to $+3.1$~eV) and not attractive. It is the only structure in the set in which electrostatics raises the barrier and adds to the steric wall. Tetragonal \ce{Li7La3Zr2O12} is the only case in which the electrostatic term is a full contribution and not a small correction. At the saddle $E_{\mathrm{elec}}=-3.99$~eV, comparable in magnitude to the induced-dipole contribution of $-4.26$~eV. This reflects the ordered, fully occupied lithium sublattice and the high-valence framework. 

The decomposition therefore transfers across transport networks that differ in chemistry and dimensionality. From a single static structure, the method identifies the physical origin of each transport bottleneck without empirical calibration. Any mechanistic descriptor must have this property if it is to be used for prediction beyond the validation set used for benchmarking.

\subsection{Route, bottleneck, and dimensionality}
\begin{figure}[htbp]
    \centering
    \includegraphics[width=0.8\textwidth]{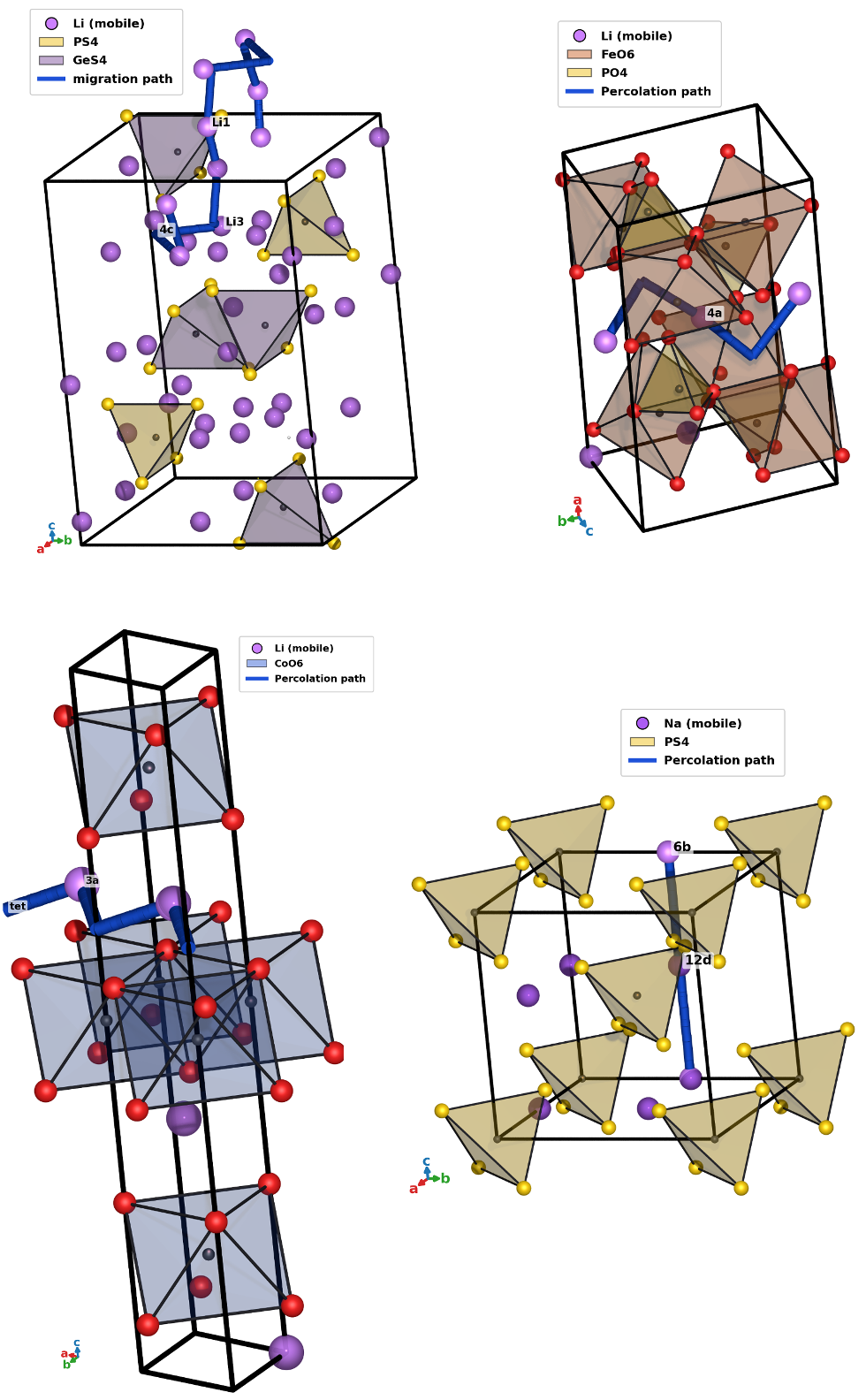}
    \caption{Predicted ionic migration networks and routes for four representative structures.}
    \label{fig:ladder}
\end{figure}

The barrier-threshold topology gives the correct route, bottleneck, and dimensionality for every structure, in agreement with the mechanisms found in independent experimental and computational studies in the literature. Figure~\ref{fig:ladder} shows the predicted networks
and routes for four representative structures, and Table~\ref{tab:summary} compares the results for the full set with the references.

We take \ce{Li10GeP2S12} as the representative model for three-dimensional transport. Early studies of long-range diffusion described the transport as essentially three-dimensional and isotropic, based both on neutron
diffraction with bond-valence and nuclear-density analysis and on measurements that average over the meandering paths between channels \cite{Weber2016}. Later impedance spectroscopy on single crystals resolved a weak but definite anisotropy. The conductivity along the $c$ axis is about four times larger than in the $ab$ plane, while the activation energies in the two directions are nearly equal, so long-range transport remains three-dimensional \cite{Iwasaki2019}. TSPD reproduces this behavior from the static structure alone. With the parameter-free energetics, all three axes open at the same threshold, with an ordinal anisotropy of $0.000$~eV. This reproduces both the three-dimensional character and the near equality of the measured activation energies. The remaining factor of four in the conductivity ratio has no counterpart in the static barriers and is not recovered. Since the activation energies are isotropic, this ratio originates in the transport prefactor and not in the migration barrier. It is a rate quantity and lies outside the scope of a mechanism descriptor. This case separates the two levels clearly: the framework fixes an isotropic three-dimensional mechanism, and the dynamics fix a mildly anisotropic rate. The continuous migration pathway, which we verified structurally with an spglib symmetry analysis (SI Section~S12, Table~S4), runs along two crystallographically distinct $16h$ lithium positions (the \ce{Li1} and \ce{Li3} channel sites), which are cross-linked through an interstitial $4c$ (\ce{Li4}) site. The rate-limiting bottleneck is the intra-channel \ce{Li1}$\to$\ce{Li3} hop, which is coordinated by two framework sulfur atoms at $2.32$--$2.33$~\AA{}, with Ge in the second shell (SI Table~S2). This constriction is the structural origin of the diffusion bottleneck that narrows upon Sn substitution, as reported in earlier studies of the inductive effect.\citep{Krauskopf2018} The site-by-site route confirms all of these results (Figure~\ref{fig:lgps}a). 

\ce{LiFePO4}, in contrast, is the representative case of one-dimensional transport. For this olivine framework, TSPD finds an open transport pathway only along the $b$ axis, while transport along the $a$ and $c$ directions is blocked by high structural barriers (SI Table~S1: open axis $b$ only; SI Table~S2). The resulting curved migration path through the coordinating oxygen windows reproduces the curved $[010]$ probability density of the mobile ion obtained earlier from neutron diffraction and maximum-entropy method (MEM) analysis of partially delithiated \ce{Li}$_{0.6}$\ce{FePO4}.\citep{Nishimura2008} When this static pathway is overlaid on the host crystal structure, the parameter-free topological boundaries closely follow the experimentally observed nuclear density maps.
The remaining structures follow the ladder systematically. \ce{LiCoO2} shows two-dimensional
transport in the $ab$ plane, with the $c$ direction blocked by the \ce{CoO2} slab. The void-refined route passes
through the tetrahedral interstitial instead of going directly between octahedra, with a saddle of
$7.19$~eV compared with $10.7$~eV for the direct hop. The tetrahedrally coordinated activated state
agrees with the migration intermediate identified by Van der Ven and Ceder.\citep{VanderVen2000,VanderVen2001}
For cubic \ce{Na3PS4} we obtain fully isotropic three-dimensional transport, with the saddle on the
$12d$ Wyckoff site,  two neighboring sulfur atoms at $2.63$~\AA{}, and the resting \ce{Na} on
$6b$ (SI Table~S2). The resulting route $6b\!\to\!12d\!\to\!6b$ agrees with the established assignment, in which the
partially occupied $12d$ interstitial is the transition point.\citep{Na3PS4ref} Tetragonal \ce{Na3PS4} has the same composition but lower symmetry and therefore isolates the structural origin of the anisotropy. The $a$ and $b$ axes open at threshold, while $c$ opens at a barrier $2.34$~eV higher. At accessible barriers the phase is thus effectively two-dimensional. The low-barrier axes are curved and interstitial-refined, and the high-barrier axis is straight (SI Table~S1). \ce{Li6PS5Cl} has the most open transport network according to the intensive descriptors. It is a fully isotropic three-dimensional network whose route winds over $26$~\AA{} through inter-cage windows. The resting sites lie on the $24g$ (\ce{T5a}) orbit, and the rate-limiting hop is surrounded by three sulfur atoms, with a chlorine at $3.3$~\AA{} in the second shell. This chlorine position is consistent with the finding that the halogen distribution controls the inter-cage jump frequency (SI Tables~S2 and~S4).\citep{Baktash2020}

The two ordered oxides at the end of the ladder separate geometric connectivity from migration.
In tetragonal \ce{Li7La3Zr2O12}, tetrahedral $8a$ and octahedral $32g$ lithium sites
alternate through an oxygen window at $1.76$~\AA{} (SI Tables~S2 and~S3). The geometric verdict is three-dimensional, but the
fully occupied lithium sublattice leaves no site for a single-ion hop, so the structure is
occupancy-walled. This is the static counterpart of the established result that tetragonal LLZO requires
concerted migration, whereas cubic LLZO conducts by single hops.\citep{Meier2014,He2017}
$\gamma$-\ce{LiAlO2} shows the separation most clearly. Its lithium sublattice is geometrically
connected in all three directions, with $(40,40,58)$ routes along the three axes. Migration along $a$ and $b$ is nevertheless blocked by a wall of oxygen and aluminum, while $c$ conducts at $1.80$~eV. This recovers the one-dimensional $c$ channel established by tracer, neutron, and NMR measurements.\citep{Wiedemann2016} Geometrically open positions along $a$ and $b$ therefore do not form a transport pathway. Only the energetics of the wall distinguish a real three-dimensional sublattice from a one-dimensional transport network whose mobile sites happen to be neighbors across a wall. This is the decisive test of the mechanism concept. TSPD passes it because the dimensionality is determined from the energetics across the wall and not from geometric adjacency.

\begin{table*}[t]
\centering
\caption{Transport-mechanism descriptors and validation for the eight structures. The conduction
dimensionality is the number of axes that open under the energetics. The route and bottleneck are
taken from the serialized minimax path. The framework shell lists the atoms bordering the saddle.
$\lambda_2^{\mathrm{mob}}$ is the algebraic connectivity of the resting-site network and measures the network density. The vacancy verdict is obtained from the occupancy-aware response. Barriers are ordinal. The complete set of numerical values, including $B_{\mathrm{perc}}$, the ordinal anisotropy, $B^\ast$, $\Delta B_1$, and the degeneracy, is given in SI Table~S1.}
\label{tab:summary}
\small
\setlength{\tabcolsep}{4pt}
\begin{tabularx}{\textwidth}{@{}l c Y c Y Y Y@{}}
\toprule
Structure & Dim. & Route / bottleneck & $\lambda_2^{\mathrm{mob}}$ & Framework shell (\AA) & Vacancy verdict & Reference \\
\midrule
\ce{LiFePO4} & 1D & curved [010] \ce{Li}--\ce{Li} & 2.00 & O 1.74, O 2.41 & opens $b$, $c_c\!\approx\!0.375$ & neutron \citep{Nishimura2008} \\
\ce{LiCoO2} & 2D & in-plane via tetrahedral interstitial & 4.60 & O 1.73, O 1.73 & opens $ab$, $c_c\!\approx\!0.125$ & CE+kMC\citep{VanderVen2000,VanderVen2001} \\
\ce{Li10GeP2S12} & 3D & \ce{Li1}$\to$\ce{Li3} via $4c$ & 2.00 & S 2.32, S 2.33 & 3D at operating comp. & neutron/AIMD \citep{Weber2016,Krauskopf2018} \\
\ce{Na3PS4} (cub.) & 3D & $6b\!\to\!12d\!\to\!6b$ & 1.17 & S 2.63, S 2.63 & vacancy-mandatory, 3D & AIMD\citep{Na3PS4ref} \\
\ce{Na3PS4} (tet.) & 3D$^{a}$ & \ce{Na}--\ce{Na} ($c$ hardest) & 1.37 & S 2.57, S 2.57 & vacancy-mandatory & symmetry split \\
\ce{Li6PS5Cl} & 3D & $24g$ inter-cage window & 6.00 & S 2.25--2.68, Cl 3.30 & vacancy/disorder & AIMD\citep{Baktash2020} \\
tet-\ce{Li7La3Zr2O12} & 3D$^{b}$ & $8a\!\to\!32g$ via O window & 9.84 & O 1.76, O 1.83 & walled at full occ. & AIMD\citep{Meier2014,He2017} \\
$\gamma$-\ce{LiAlO2} & 1D & [001] \ce{Li}--\ce{Li} & 2.00 & O 1.71, O 1.78 & $c$ opens under vac. & neutron\citep{Wiedemann2016} \\
\bottomrule
\end{tabularx}

\vspace{2pt}
\footnotesize $^{a}$Effectively two-dimensional at accessible barriers ($c$ opens $2.34$~eV higher).\quad
$^{b}$Geometric connectivity is three-dimensional; occupancy-walled at the real composition.
\end{table*}



\subsection{Redundancy of the conduction network}
Here we evaluate the redundancy of the conduction network and the degeneracy of the rate-limiting window. Both descriptors are needed because they describe different modes of failure. Redundancy is the availability of an alternative route once the rate-limiting window is lost. This window is not unique. It occurs as a set of crystallographically equivalent copies with identical barriers. We remove the whole set at once and not one copy at a time, because a substitutional dopant occupies a crystallographic site and therefore removes all copies simultaneously. The minimax percolation is then solved again on the remaining network, and $\Delta B_1$ is the additional barrier that the mobile ion must overcome to use the surviving route (SI Section~S4). The degeneracy is the symmetry multiplicity of the rate-limiting saddle, i.e., the number of crystallographically identical copies of the window, which have identical barriers by symmetry. The values for each structure are listed in SI Table~S1.

Comparing the direct-hop barrier $B^\ast$ with the percolation saddle distinguishes
direct channel transport from interstitial-mediated transport. In the direct case,
for example \ce{LiFePO4} ($B^\ast=0.035$~eV), \ce{Li10GeP2S12} ($0.567$~eV),
tetragonal \ce{Na3PS4} ($1.090$~eV), and $\gamma$-\ce{LiAlO2} ($0.150$~eV), $B^\ast$ lies
below the percolation saddle. The easiest hop between two mobile-ion sites then has a low barrier compared with the network bottleneck, and transport proceeds along the direct channel. In the interstitial case, $B^\ast$ lies above the saddle, so direct migration is unfavorable and transport has to go through intermediate sites. In \ce{LiCoO2}, $B^\ast$ lies $3.48$~eV above the saddle, consistent with the divacancy mechanism through a transient tetrahedral site. The difference is $0.97$~eV for tetragonal \ce{Li7La3Zr2O12} and $7.71$~eV for \ce{Li6PS5Cl}. The latter is the largest in the set, so every conducting event in the argyrodite is a hop through an inter-cage window and not a direct hop. Cubic \ce{Na3PS4} is the boundary case. Here $B^\ast$ equals $B_{\mathrm{perc}}$ to within a meV, so the direct hop and the percolation hop coincide. The signed difference $B^\ast-B_{\mathrm{perc}}$ is given for all eight structures in SI Table~S1.

\begin{figure}[htbp]
    \centering
    \includegraphics[width=\textwidth,height=0.40\textheight,keepaspectratio]{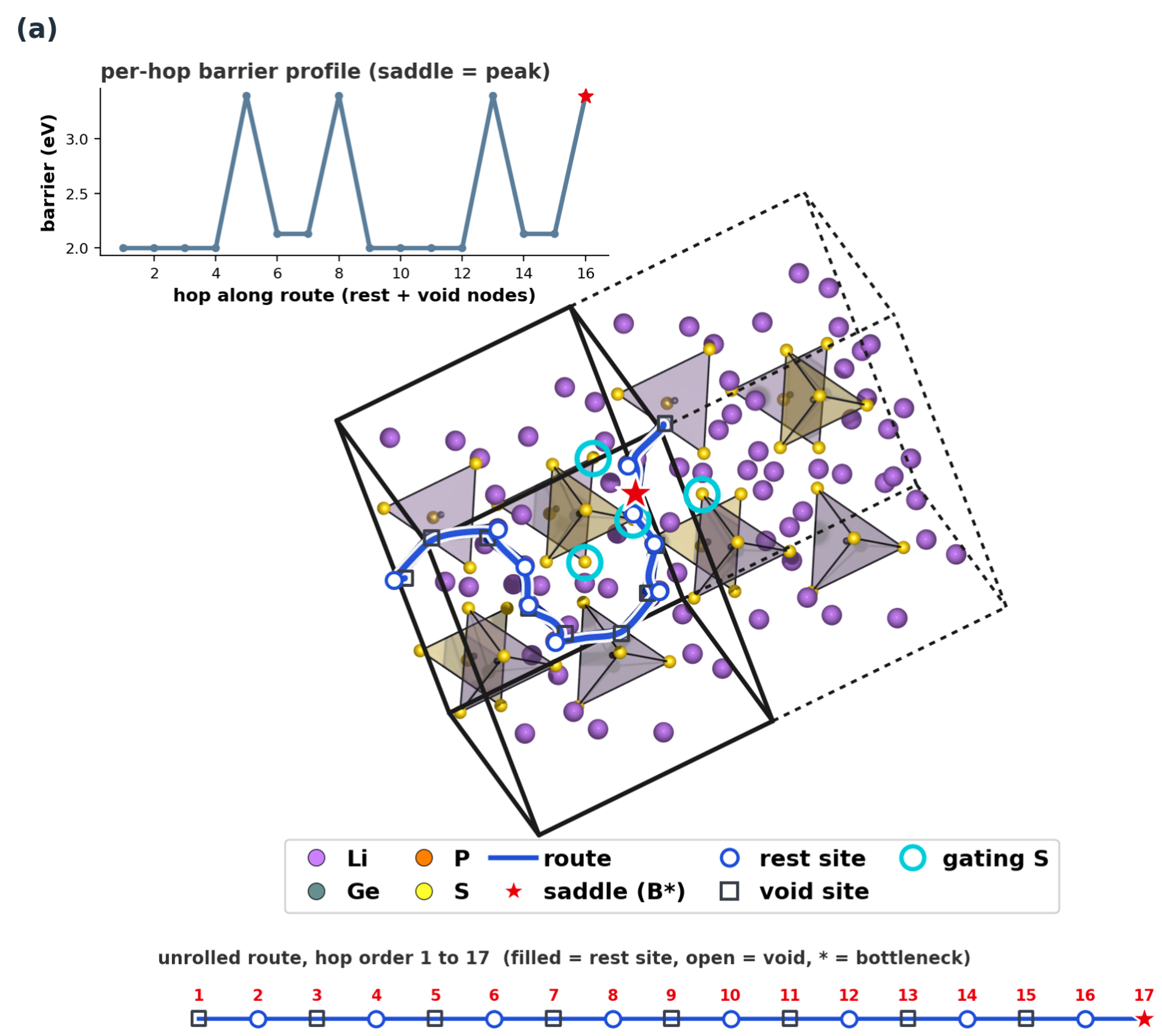}\\[6pt]
    \includegraphics[width=\textwidth,height=0.40\textheight,keepaspectratio]{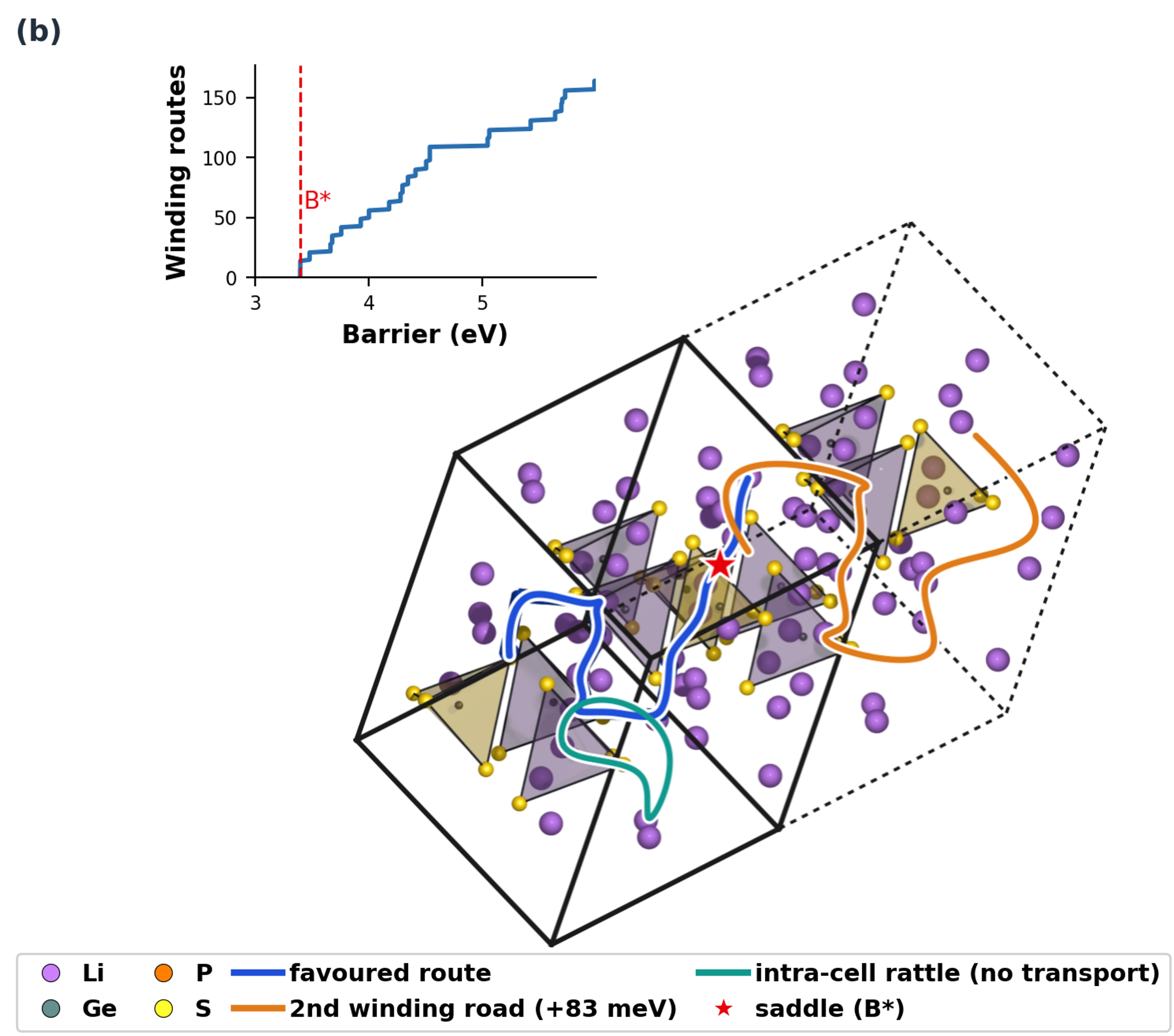}
    \caption{Bottleneck and redundancy in \ce{Li10GeP2S12}. (a) Rate-limiting crossing on the winding route. The migration-barrier profile of each hop (inset) has its maximum at the saddle $B^\ast$. The route alternates between resting \ce{Li} sites and interstitial voids, and the bottleneck is the \ce{Li1}$\to$\ce{Li3} channel-to-channel hop bordered by two framework sulfur atoms. The bottleneck position is marked on the unrolled hop sequence. (b) Redundancy. The preferred route and a second winding path only $83$~meV higher both span the cell, whereas the motion within a cage does not contribute to long-range transport.}
    \label{fig:lgps}
\end{figure}

The redundancy dimensionality divides the set unambiguously. \ce{LiFePO4} and
$\gamma$-\ce{LiAlO2} are one-dimensional. Every independent route runs along a single
axis, so an obstruction cannot be relieved by rerouting into another direction. This is
the static structural counterpart of the experimental observation that a single anti-site
defect blocks an entire olivine channel.\citep{Malik2010,Morgan2004,Islam2005}
\ce{LiCoO2} is two-dimensional, and \ce{Li10GeP2S12}, both \ce{Na3PS4} polymorphs, and
\ce{Li6PS5Cl} are three-dimensional. In these networks a blocked bottleneck can be bypassed along an independent direction, and the ablation sequence gives the energy cost of this alternative. \ce{Li10GeP2S12} is the representative redundant case
(Figure~\ref{fig:lgps}b). Once the orbit of its rate-limiting hop is blocked, the next
percolating route lies only $83$~meV above the primary route (SI Section~S4). The
one-dimensional structures are at the opposite limit, with $296$~meV for \ce{LiFePO4} and
$1.58$~eV for $\gamma$-\ce{LiAlO2}. The two \ce{Na3PS4} polymorphs bracket the effect of
the tetragonal distortion. The cubic polymorph reroutes at a cost of $48$~meV and the tetragonal polymorph at a cost of $275$~meV.

Degeneracy and redundancy describe different failure modes, and this difference is why both are needed. A point defect removes a single atom and therefore a single window.
Degeneracy counts the crystallographically identical copies of the rate-limiting
window. These copies have the same energy by symmetry, so an equivalent copy can take over the transport with no additional barrier. A substitutional dopant, however, occupies a
crystallographic site and therefore removes all symmetry-equivalent copies at once. In that case degeneracy gives no protection, because the set it counts
is the set that is removed. This is the case that $\Delta B_1$ addresses, since it is obtained by removing the whole orbit and solving the minimax percolation again on the remaining network.

Degeneracy alone does not give fault tolerance and must be considered together
with the conduction dimensionality. \ce{LiFePO4} has $64$ copies of its
rate-limiting window, the second-largest number in the set, yet a single anti-site
defect blocks an entire channel.\citep{Malik2010,Morgan2004,Islam2005} Because the structure is one-dimensional, the $64$ copies are distributed over $64$ independent channels and do not form alternative routes within one channel. A blocked
window in one channel therefore cannot be replaced by an identical window in another. In
\ce{Li6PS5Cl}, a degeneracy of the same order does give real tolerance, because the
network is three-dimensional and the copies can be reached from one another. Tolerance to point defects therefore requires both a large degeneracy and a conduction
dimensionality of at least two, and neither is sufficient alone.

The redundancy
dimensionality and the number of bottleneck types are intensive quantities that do not depend on the choice of cell. They can therefore be used across chemistries and transport dimensionalities as criteria for the fault tolerance of a conductor. The network degeneracy, in contrast, is the site-symmetry multiplicity of the bottleneck environment. Cubic \ce{Na3PS4} has a very uniform profile, with many symmetry-equivalent windows and a barrier spread of only $0.000$~eV, which is characteristic of a high-symmetry superionic framework. In $\gamma$-\ce{LiAlO2} the spread across the equivalent parallel passes is also essentially zero. The interstitial-mediated cage networks have the broadest spread, reaching $0.199$~eV in \ce{Li6PS5Cl}, which reflects the less uniform inter-cage windows.

The algebraic connectivity $\lambda_2$ (SI Table~S1) is the continuous spectral counterpart of the minimum cut that underlies the redundancy dimensionality, and it gives a lower bound on this cut through the Cheeger inequality. It also measures the network density and not the migration dimensionality. It is highest for the dense sublattice of tetragonal \ce{Li7La3Zr2O12}, even though this network is occupancy-walled. We do not report a separate comparison of the homological and spectral descriptions of the bottleneck, because the two coincide by construction. The only
question such a comparison could answer is whether the rate-limiting hop lies on a cycle that can be bypassed, and this question
is already answered by the ordering of $B^\ast$ and the saddle, by the ablation sequence of alternative routes, and by $\lambda_2$. This point applies to the whole analysis. The redundancy descriptors characterize the
transport network geometrically and must be read together with the occupancy verdict, because a geometrically redundant sublattice can still be occupancy-walled at its real composition.

\subsection{Occupancy and the mechanism--rate separation}
\begin{figure}[htbp]
    \centering
    \includegraphics[width=\textwidth,height=0.40\textheight,keepaspectratio]{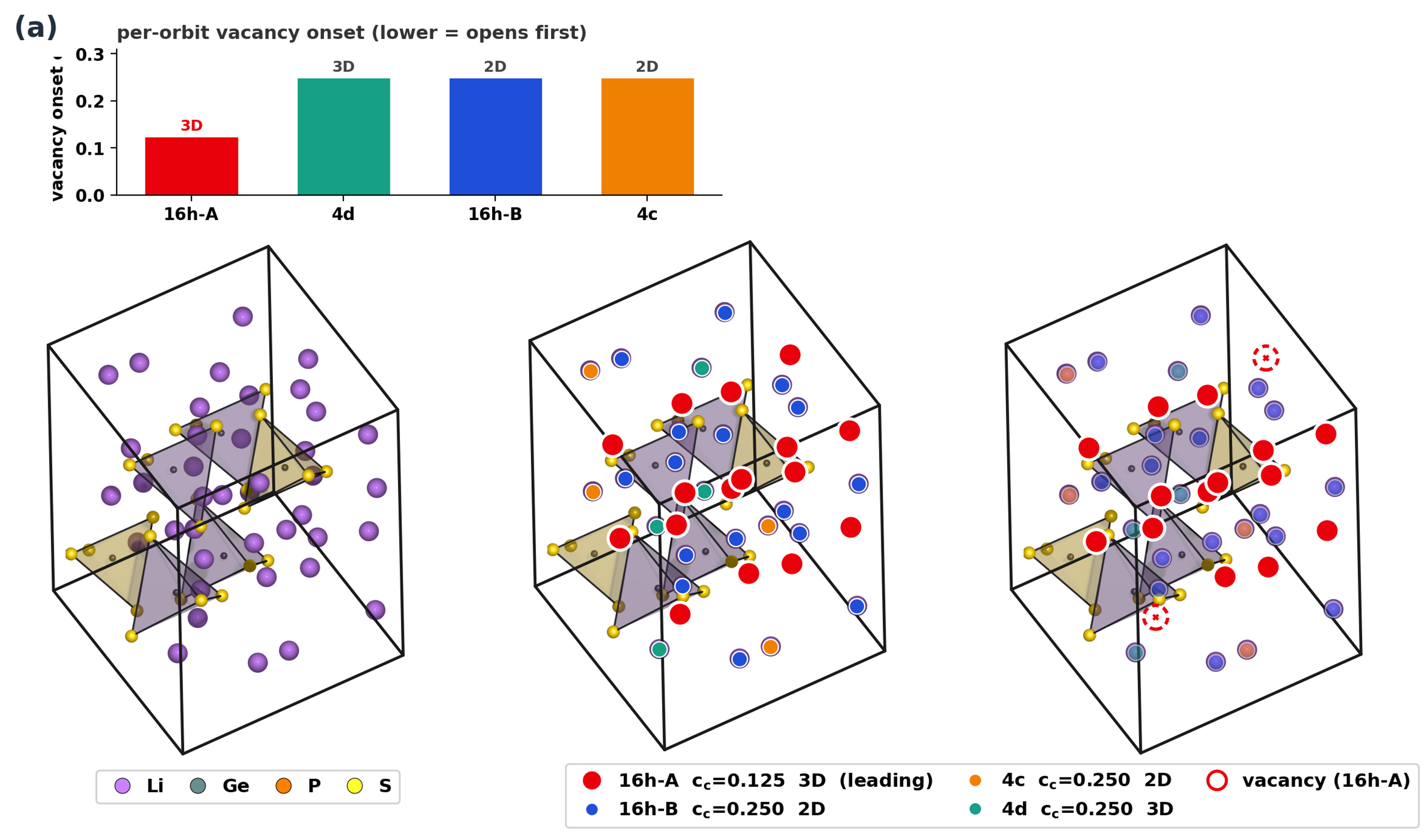}\\[6pt]
    \includegraphics[width=\textwidth,height=0.40\textheight,keepaspectratio]{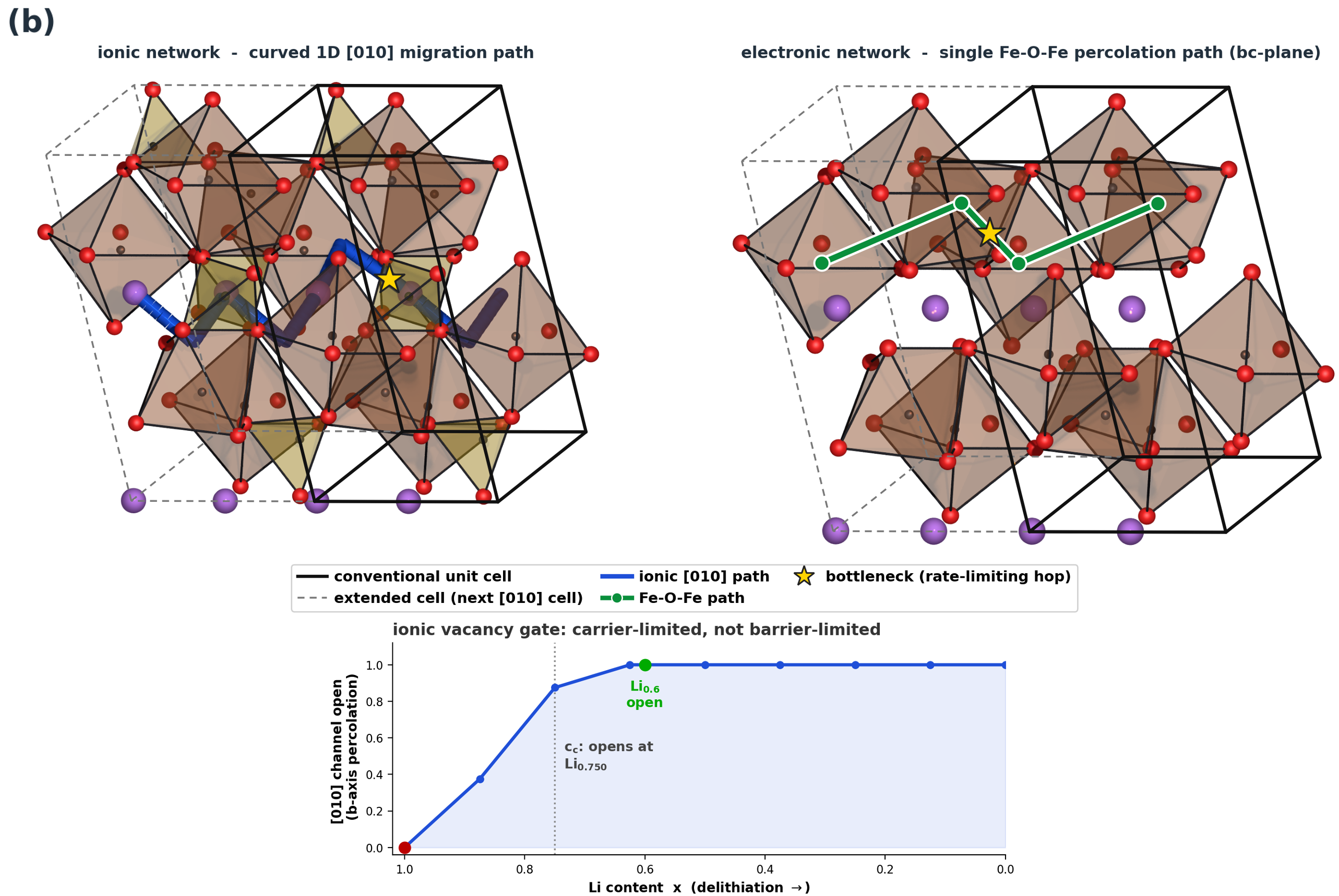}
\caption{Occupancy and the mechanism--rate separation. (a) Vacancy response of each site class in \ce{Li10GeP2S12}. Vacating the leading $16h$ lithium class opens three-dimensional percolation
first, at $c_c=0.125$, before the other three symmetry-inequivalent classes. (b) Dual network
in \ce{LiFePO4}, consisting of a one-dimensional ionic [010] channel and a three-dimensional
\ce{Fe}--O--\ce{Fe} electronic network. The ionic vacancy gate is limited by carriers and not by the barrier, and the fixed [010] neck does not change with lithium content.}
\label{fig:vac}
\end{figure}

Including the site occupancy converts the static geometric mechanism into a description of the actual stoichiometric composition. This step explicitly separates the structural transport mechanism from the macroscopic kinetic rate. Figure~\ref{fig:vac}a shows the site-resolved vacancy response of \ce{Li10GeP2S12}, which we chose as a model system because it has four symmetry-inequivalent lithium sites.
This allows us to determine which Wyckoff site class starts dimensional percolation when vacancies are introduced, and at which critical threshold. Vacancies in the leading $16h$ class open three-dimensional percolation first, at a
vacancy fraction of $c_c=0.125$, before any of the other three classes. The analysis thus identifies the specific crystallographic vacancy that unblocks the network. The occupancy-blind topological results then give explicit, composition-dependent criteria for transport. At full stoichiometric occupancy, the cubic and tetragonal \ce{Na3PS4} phases and
tetragonal \ce{Li7La3Zr2O12} have no percolating pathway, so long-range transport in these materials depends entirely on defects. This agrees with the experimental observation that pristine cubic \ce{Na3PS4} is a poor conductor whose ionic conductivity depends strongly on sodium excess or aliovalent doping.\citep{Na3PS4ref} Similarly, the intercalation cathodes require some initial delithiation before they percolate, and \ce{LiFePO4} becomes active at a critical vacancy fraction of $c_c\approx0.375$, close to the degree of delithiation at which the experimental
path was imaged. 

The vacancy response completes the mechanistic assignments begun earlier in this section. For \ce{LiCoO2}, the ordering $B^\ast>B_{\mathrm{perc}}$ showed that the
percolating route cannot consist of direct octahedral hops and must pass through an
intermediate. The serialized route identified this intermediate as the tetrahedral site
bordered by oxygen. The occupancy-aware response gives the carrier content at
which the route becomes available, i.e., the condition under which the tetrahedral
intermediate is accessible to a migrating ion and not blocked by an occupied neighbor. The three descriptors are independent, and together they reproduce the
divacancy mechanism of the layered oxide from the static structure
alone.\citep{VanderVen2000,VanderVen2001} The same argument applies to \ce{Li6PS5Cl}, where $B^\ast$ exceeds $B_{\mathrm{perc}}$ by $7.71$~eV, the largest
difference in the set. No direct \ce{Li}--\ce{Li} hop is accessible below $7.8$~eV,
while the network percolates at $0.10$~eV. Every conducting event is therefore mediated, and
the serialized route identifies the mediating site as the inter-cage window bordered by
sulfur. In neither material is the mechanism inferred from the geometry of the route. It is
required by the ordering of two barriers and then resolved by the serialized route and the occupancy analysis.

Tetragonal \ce{Li7La3Zr2O12} shows this separation most clearly. Its dense, fully occupied, occupancy-walled sublattice allows no single-ion trajectory, so the geometric openness of the static structure does not translate into single-ion percolation at all. In this regime, the static framework sets the structural constraints under which transport can occur only through strongly correlated, collective mechanisms.

We also applied the graph-theoretic framework to a second network in cathode materials in which polaronic hopping occurs. Figure~\ref{fig:vac}b shows this for \ce{LiFePO4}, where the
electronic network is formed by \ce{Fe}--O--\ce{Fe} superexchange links. The ionic network is the
one-dimensional [010] channel, while the electronic network percolates in three dimensions. The minimum cuts of the ionic and electronic networks lie in different regions of the structure. \ce{LiFePO4} is therefore a mixed
conductor whose ionic and electronic limitations are decoupled, so that each can be optimized through separate structural or compositional parameters. The vacancy
gate of the ionic channel is limited by carriers and not by the barrier. The fixed [010] bottleneck does not change with lithium concentration, and long-range transport is switched on only by adding charge carriers. Across the
set, the dual-network classification distinguishes intercalation cathodes, which are mixed conductors with spatially separate ionic and electronic bottlenecks, from sulfide solid electrolytes, which are ideal ionic conductors because they have no electronic network mediated by transition metals.

\section{Discussion}
TSPD gives the transport mechanism to the extent that it can be determined from a single static structure. From the crystal framework it obtains the path and dimensionality of the transport pathways, the rate-limiting bottlenecks together with the framework atoms that coordinate them, the topological redundancy of the percolating network, and the critical vacancy fractions needed to start long-range transport. The descriptors are parameter-free and calibration-free, and the predictions are validated against migration-ion densities from AIMD and against neutron diffraction data. Our claims are therefore kinematic and mechanistic, not kinetic. They require only a host structure and a mobile species, at a cost of seconds per structure.

The design of the method implies several limitations, which we state here as properties of the method. First, the energy scale is ordinal. The barrier of each edge ranks crystallographic directions and locates the bottleneck within a structure, but it is not a transport rate and does not give the absolute separation between competing directions. Second, the method resolves the mechanism and not the rate. It predicts neither a conductivity nor a diffusivity, because carrier concentration, temperature, and correlated motion are outside a static treatment by construction. Third, strongly cooperative motion is not modeled, although its static precondition is often visible. Tetragonal \ce{Li7La3Zr2O12} is the clearest example: its dense, fully occupied, occupancy-walled sublattice offers no single-ion path, and conduction occurs only collectively. When a reference mechanism has a correlated or concentration-dependent component, TSPD claims only the geometry and the activated site that it actually reproduces and assigns the rest to the rate. \ce{LiCoO2} is the model case. The route through the tetrahedral interstitial and the corresponding activated state are recovered, whereas the divacancy character of the low-barrier process and its dependence on lithium concentration and cobalt valence are configurational and dynamical effects on the rate, outside a single static treatment. Fourth, the input is a single ordered structure. Site disorder and partial occupancy enter through the vacancy response and not as a configurational average. For the argyrodite, the open geometric pathways of the ordered model are therefore reported correctly, but the dependence of the rate on order and disorder is not. Finally, we report redundancy through the edge-disjoint route ratio and $\lambda_2$ and not through a comparison of the homological and minimax bottlenecks. On a barrier filtration the two coincide by
construction, and their comparison carries no information beyond the
$B^\ast$-to-saddle ordering already reported (SI Section~S6).

Taken over the whole benchmark set, the analysis changes the question that screening asks. A measured or predicted conductivity tells us whether a material conducts. The mechanism map tells us how it conducts: in which directions, through which atoms, at what carrier content, and with what tolerance to obstruction. This is possible because the topological and spectral descriptors are obtained independently of the carrier population. A host framework that has open pathways and is ready to conduct once carriers are inserted into individual sites can therefore be distinguished from a lattice that is intrinsically insulating. This distinction cannot be made by analyzing the occupied cell alone. An apparent absence of macroscopic transport can then be attributed to the carrier composition and not to the connectivity of the network. The vacancy response turns this into a design map. It identifies the Wyckoff sites whose vacancies open a given transport dimensionality and the vacancy fraction at which this happens. This is the information needed for a doping or off-stoichiometry strategy, and it is obtained before any defect calculation. The redundancy and degeneracy descriptors add an independent dimension, the fault tolerance of a conductor. They distinguish transport that relies on a single fragile route, which one isolated defect can block, from transport spread over independent routes, which survives the disorder present in a real material. None of these quantities can be obtained from the transport rate. Each is frame invariant, is obtained from the static structure, and can be acted on directly.

The same premise determines the outlook. The method needs only a static framework and a mobile guest, so it can be applied to transport through other structured hosts, such as porous sorbents, ion-exchange membranes, and biological channels, whenever access is controlled by a static scaffold and not by the dynamics of the species that populate it. In solid-state ionics, the mechanism map is fast enough to rank large libraries of candidates. Because it gives a falsifiable mechanism and not a fitted number, it can serve as a mechanistic pre-filter that directs expensive dynamical simulations to the structures, and to the specific hops, that deserve them. Used in this way, it complements rate-based screening and does not compete with it. The mechanism, the vacancy map, and the fault-tolerance descriptors are all defined by the structure and vary continuously with composition. They are therefore natural objectives for inverse design, in which one looks for a structure with not just an open route but a redundant route that opens at an accessible carrier content. They also provide a basis for relating short-range order and phase changes to transport. The present work establishes and validates the mechanism itself; applying these descriptors to the purposes above is left for future work.

\section{Conclusions}
A single static structure already contains a mechanism-level map of where its ions can move. We have shown that this map can be obtained in two steps: constructing a migration network weighted by physics-based energies, and analyzing it with topological and spectral methods. From one structure, the method gives the dominant route, its dimensionality and anisotropy, the rate-limiting bottleneck and the neighboring framework atoms, the redundancy and degeneracy of the available percolation paths, and the vacancy fraction at which long-range transport starts. The edge barriers are based on parameter-free physics, and the analysis combines the spectral, homological, and minimax-path descriptions of the same network.

We validated the method by comparing the predicted pathways directly with AIMD and
neutron diffraction data for eight conductors with one- to three-dimensional transport. The predicted transport dimensionality, route, bottleneck, framework atoms, and vacancy requirement agree with the reference mechanisms. Examples are the curved one-dimensional [010] channel of \ce{LiFePO4}, the two-dimensional plane of \ce{LiCoO2}, the $12d$ saddle of cubic \ce{Na3PS4}, the inter-cage window of \ce{Li6PS5Cl} framed by sulfur and chlorine, and the \ce{Li1}$\to$\ce{Li3} hop of
\ce{Li10GeP2S12}. The ordered oxides are correctly identified as having no single-ion path. The only disagreements concern the magnitude of a nearly isotropic anisotropy and dynamical effects that lie outside a static treatment by design.

The approach therefore gives a falsifiable, mechanism-level map of ion accessibility together with parameter-free energy profiles, at negligible computational cost. It moves high-throughput materials screening from the empirical identification of fast ion conductors toward a deterministic account of the transport physics behind them. It also establishes the structural transport mechanism as a distinct property, decoupled from the rate, which provides a firm basis for future predictive workflows and generative materials design.

\section*{Acknowledgements}
The authors acknowledge financial support from the Independent Research Foundation Denmark with grant no. 3164-00297B (ADANA), 2035-00232B (TeraBatt); the Novo Nordisk Foundation with grant number NNF25OC0101622(AutoMLP)

\section*{Supporting Information}
The Supporting Information contains the definitions of the network construction; the Born--Mayer steric, screened-Ewald, induced-dipole, and softness contributions to the edge energetics; the three objectives; the dimensionality, bottleneck, redundancy-dimensionality, secondary-route barrier increment, degeneracy, vacancy, saddle--cut, and dual-network descriptors; the graph-Laplacian spectrum, persistent homology, and Bloch electronic spectrum; and the route-serialization and frame-annotation procedure. It also contains the descriptor tables, serialized routes, and complete figure set for each structure, together with the datasets and computational details.

\end{document}


\maketitle
\tableofcontents
\vspace{1em}

\noindent This Supporting Information collects the mathematical definitions of the network
and its edge energetics, the descriptor definitions, the spectral and topological formalisms
(graph-Laplacian spectrum, persistent homology, and the Bloch electronic spectrum), the
route-serialization and frame-annotation procedure, the datasets and computational details,
the alternative route ladder, the per-structure data tables read
directly from the network-topology outputs of the current code version, the full
per-structure figure set, and the route validation status. All energies labeled ordinal are
used only for ordering directions and anchoring within a structural family; they are ordinal
migration barriers and are not transport rates.

\section{Migration network construction}\label{si:network}
Given a structure $\mathcal{S}$ and a mobile species $m$, the candidate mobile sites are the
crystallographic sites whose composition contains $m$. A species- and bonding-aware filter
excludes any cation bonded into an anion polyhedron, independent of its fractional occupancy,
so that framework cations (for example the \ce{Ge}/\ce{P} of a thiophosphate or the
\ce{Fe}/\ce{P} of an olivine) never enter the graph. The mobile sites form the pristine
resting-site graph $G$. The augmented network $\hat G$ adds interstitial void nodes from a
Voronoi seeding of the framework and the geometrically admissible hops between neighboring
nodes. Each node $n$ carries a fractional coordinate $f_n$ and a Cartesian coordinate $c_n$
in the analysis frame (Section~\ref{si:frame}). Edges $\{u,v\}$
are stored with the periodic image $\mathbf{j}_{uv}$ that realizes the hop, so routes can be
unwrapped across the cell boundary.

Two design choices make the connectivity verdicts origin-invariant. The Voronoi seeder
returns an identical void set under a shift of the cell origin, in contrast to a grid seeder,
whose maxima move with the origin. A void is retained to decorate a hop only where the direct
resting-site neck is already passable, and void--void chaining is forbidden, so voids annotate
already-open routes rather than create connectivity. Connectivity and dimensionality are
therefore analyzed from the resting-site sublattice, a crystallographic and origin-invariant
object, while the void-augmented network supplies the full curved route.

\section{Parameter-free edge energetics}\label{si:energetics}
The migration energy along an edge is sampled at $N$ points $P(t)=(1-t)\,c_u + t\,c_v$,
$t\in[0,1]$, and is the sum of three parameter-free terms.

\paragraph{Signed Born--Mayer steric term.}
For a probe position $P$ the steric energy is summed over framework neighbors $j$ on the true
ion--framework distance $d_j=\lVert P - c_j\rVert$,
\begin{equation}
E_{\mathrm{ster}}(P)=\sum_{j\in\mathrm{fw}} A_j\,
\exp\!\left(-\frac{d_j-(r_m+r_j)}{\rho_j}\right)\,\mathrm{sgn}\!\left(r_m+r_j-d_j\right),
\label{eq:steric}
\end{equation}
where $r_m,r_j$ are the Shannon radii of the mobile ion and neighbor and $\rho_j$ is the
Born--Mayer range. The sign makes the term a clearance measure: positive (repulsive) inside
contact, negative outside. The prefactor $A_j$ is fixed by force balance against the
electrostatic term at the equilibrium coordination,
\begin{equation}
\left.\frac{\partial E_{\mathrm{ster}}}{\partial d}\right|_{d=d^{\mathrm{eq}}}
+ \left.\frac{\partial E_{\mathrm{elec}}}{\partial d}\right|_{d=d^{\mathrm{eq}}} = 0,
\label{eq:forcebalance}
\end{equation}
so no constant in Eq.~\eqref{eq:steric} is fitted to a diffusivity or a migration barrier.

\paragraph{Anion softness (HSAB).}
The Born--Mayer range is scaled by an anion-softness factor $s_j$ from hard--soft acid--base
chemistry,
\begin{equation}
\rho_j=\rho\,s_j^{1/2},
\label{eq:softness}
\end{equation}
so polarizable anions (sulfide, chloride) present a softer, longer-range wall than hard
oxides. The exponent $1/2$ is the single documented modeling assumption and is togglable for
ablation.

\paragraph{Screened-Ewald electrostatics.}
The interaction of the mobile ion (charge $z_m$) with the framework charges is
\begin{equation}
E_{\mathrm{elec}}(P)=\frac{z_m\,\phi_{\mathrm{Ewald}}(P)\,k_{\mathrm{C}}}{\varepsilon},
\label{eq:elec}
\end{equation}
where $\phi_{\mathrm{Ewald}}$ is the Ewald potential of the framework charge distribution
(mobile species removed), $k_{\mathrm{C}}$ is the Coulomb constant in \si{eV.\angstrom}, and
$\varepsilon$ is the static dielectric from a Clausius--Mossotti relation over the framework
polarizabilities.

\paragraph{Induced-dipole term.}
A polarization term $E_{\mathrm{ind}}(P)$ adds the response of framework anions of
polarizability $\alpha_j$ to the field of the mobile ion; it enters additively and softens
the barrier.

\paragraph{Per-edge barrier and decomposition.}
The site energy at a non-void node is
$E_n=E_{\mathrm{ster}}(c_n)+E_{\mathrm{elec}}(c_n)+E_{\mathrm{ind}}(c_n)$, the baseline is
the deepest mobile site $E_0=\min_n E_n$, and the edge barrier is
\begin{equation}
B_{uv}=\max_{t}\Big[E_{\mathrm{ster}}(P(t))+E_{\mathrm{elec}}(P(t))+E_{\mathrm{ind}}(P(t))\Big]-E_0,
\label{eq:edgebarrier}
\end{equation}
with the saddle (maximizing $t$) decomposed into steric and electrostatic parts. The
\emph{unified} objective uses Eq.~\eqref{eq:edgebarrier}; the \emph{geometric} objective keeps
only $E_{\mathrm{ster}}$; the \emph{electrostatic} objective keeps only $E_{\mathrm{elec}}$.
Per-structure decompositions of the rate-limiting saddle into
$E_{\mathrm{ster}}/E_{\mathrm{elec}}/E_{\mathrm{ind}}$ are extracted from the analysis outputs and
confirm that the steric term dominates the barrier in every structure of the set, with the
electrostatic term a minor correction and the induced-dipole term a softening contribution.

\section{Dimensionality, anisotropy, and the bottleneck}\label{si:dim}
For a lattice direction $a\in\{0,1,2\}$ the per-axis route is the minimax (bottleneck) path,
\begin{equation}
B_{\mathrm{perc}}^{(a)}=\min_{\pi\in\Pi_a}\ \max_{\{u,v\}\in\pi} B_{uv},
\label{eq:minimax}
\end{equation}
computed by a threshold sweep with union--find spanning detection on the resting-site
network. The material barrier is $B_{\mathrm{perc}}=\min_a B_{\mathrm{perc}}^{(a)}$. A winding
filtration raises a threshold $\beta$ and records the threshold at which each axis first acquires a
spanning route (the opening $\Delta_a$). The dimensionality is the number of axes that ever
open; the ordinal anisotropy is $\max_a\Delta_a-\min_a\Delta_a$. The bottleneck is the
rate-limiting edge of Eq.~\eqref{eq:minimax}, with its saddle fractional coordinate, symmetry
multiplicity, and framework coordination shell (the bordering species and their distances).
A companion quantity is the direct-hop barrier
\begin{equation}
B^\ast=\min\{\,B_{uv}\;:\;u,v\ \text{both resting mobile sites}\,\},
\label{eq:bstar}
\end{equation}
the easiest single mobile-to-mobile hop. It is evaluated on the pristine resting-site
graph $G$, never on the void-augmented network $\hat G$, so that an interstitial-mediated
route cannot lower it. Its position relative to $B_{\mathrm{perc}}$ is diagnostic:
$B^\ast<B_{\mathrm{perc}}$ identifies direct channel transport, in which the direct hop is
the low-barrier step and the network bottleneck lies elsewhere, whereas
$B^\ast>B_{\mathrm{perc}}$ identifies interstitial-mediated transport, in which direct
migration is penalized and the route proceeds through an intermediate site.

\section{Redundancy and degeneracy}\label{si:redundancy}
\textbf{Redundancy.} Redundancy is the availability of alternative routes within a small
barrier increment of the percolation bottleneck. It is reported as the redundancy spectrum
\begin{equation}
N(\Delta B)=\#\{\,\{u,v\}\in\hat G\;:\;B_{\mathrm{perc}}\le B_{uv}\le B_{\mathrm{perc}}+\Delta B\,\},
\label{eq:redundancy}
\end{equation}
the number of windows whose barrier lies within $\Delta B$ of the bottleneck, and by the
scalar $\Delta B_1$, the increment at which the first alternative \emph{percolating} route
appears (Section~\ref{si:redundancy}). The two are inverse readings of one curve: $N(\Delta B)$
fixes the energy and counts routes; $\Delta B_1$ fixes the route and reports the energy.
$N(\Delta B)$ counts all near-bottleneck windows, whether or not they percolate, so it
measures the local abundance of competing passes; $\Delta B_1$ counts only routes that carry
long-range transport. Both are barriers on the periodic quotient graph and are invariant
under a shift of the cell origin and a permutation of the lattice vectors.

\textbf{Degeneracy.} The symmetry multiplicity of the rate-limiting saddle, obtained by applying the space-group operations of the crystal to the saddle position and counting distinct images. Members of the orbit are related by symmetry and therefore carry identical barriers.

\textbf{The secondary-route barrier increment.}
The percolation barrier $B_{\mathrm{perc}}$ fixes the barrier of the lowest-barrier
route. What matters next is the barrier of the route that remains once that route is
obstructed. This is computed by ablation. The minimax solve of Eq.~\eqref{eq:minimax}
returns the rate-limiting hop along a given axis. That hop does not stand alone: it
belongs to a symmetry orbit of iso-energetic hops, and the ablation removes the orbit
in full rather than the single edge. The distinction is chemical. A point defect
removes one member of the orbit, and an equivalent member accepts the transport without
an increase in barrier; a substitutional dopant occupies a crystallographic site and
removes every member simultaneously. The single-edge construction accordingly returns
$\Delta B_1=0$ for every structure with a degenerate bottleneck, which is a correct
statement about point-defect tolerance but not the quantity a substitution strategy
requires.

The orbit is identified energetically, by removing every hop whose barrier lies within
$2$~meV of the rate-limiting barrier, re-solving the minimax percolation on the
remaining graph, and recording the new barrier. Iterating gives a sequence
$B_{\mathrm{perc}}=B_0<B_1<B_2<\dots$, and the increment to the first alternative route
is $\Delta B_1=B_1-B_0$. The sequence terminates when the axis is severed, at which
point the number of steps is the number of independent barrier classes the axis
carries. Because every step is a minimax barrier on the periodic quotient graph, the
sequence is invariant under a shift of the cell origin and a permutation of the lattice
vectors by the same argument that protects $B_{\mathrm{perc}}$.

Two distinct scales bear on the tolerance. Symmetry-equivalent hops are exactly
iso-energetic, so any tolerance above the numerical noise floor recovers a complete
orbit; the barrier spread across the symmetry orbit of the rate-limiting edge vanishes
to within $10^{-3}$~meV for every structure in the validation set
(Section~\ref{si:compute}). The tolerance is therefore set by the thermal scale rather
than by symmetry. At $300$~K, $k_{\mathrm{B}}T=25.9$~meV, and $2$~meV corresponds to
$0.08\,k_{\mathrm{B}}T$, so hops separated by less than the tolerance are thermally
indistinguishable and carry transport concurrently rather than in succession.
$\Delta B_1$ is accordingly the increment to the next route that is thermally distinct
from the one removed.

\ce{Li10GeP2S12} illustrates both the construction and the tolerance. Its barrier levels
are discrete, each level being one symmetry orbit: $3.3943$~eV with multiplicity $16$,
$3.3958$~eV with multiplicity $16$, and $3.4772$~eV with multiplicity $8$. The second
level lies $1.5$~meV above the first, that is $0.06\,k_{\mathrm{B}}T$, and at $300$~K
carries $48\%$ of the hopping population against the first level's $51\%$. The two
levels are not a route and an alternative but a single thermally degenerate channel of
$32$ windows, and the tolerance removes both. The next thermally distinct route is the
third level, $83$~meV or $3.2\,k_{\mathrm{B}}T$ above the bottleneck, carrying $1\%$ of
the population; the fourth follows a further $184$~meV above. The small first increment
is the static signature of a redundant conductor. Whether a substitution on a single
framework Wyckoff orbit removes both of the near-degenerate levels or only the first is
not established by the present construction, and the reported increment is the
conservative case in which the thermally degenerate channel is removed in full.

The one-dimensional structures occupy the opposite limit. \ce{LiFePO4} requires an
increment of $297$~meV to reach its first alternative route and a further $1.4$~eV to
reach the second, and $\gamma$-\ce{LiAlO2} requires $1.58$~eV to reach its first. Both
reflect the absence of a low-barrier alternative once the single channel is obstructed.
\ce{LiCoO2} is the extreme case at $3.48$~eV, consistent with a layered oxide in which
every route passes through the same oxygen-bordered window class, so that removing that
class leaves no accessible alternative at all.

The two \ce{Na3PS4} polymorphs quantify the effect of the tetragonal distortion on
fault tolerance, and the sense of the effect is opposite to that of the anisotropy. The
tetragonal phase carries the lower-barrier alternative, at $\Delta B_1=275$~meV, while
the cubic phase requires $1.62$~eV before a second percolating route becomes available.
The cubic sublattice is the more isotropic and the more highly connected of the two, yet
its routes are copies of a single window class, so that class has no low-barrier
successor. The distortion that opens an anisotropy of $2.34$~eV in the tetragonal phase
simultaneously introduces a second window class within $275$~meV of the bottleneck. The
static construction therefore separates two properties that a single conductivity
measurement reports together: the cubic phase conducts more isotropically, and the
tetragonal phase tolerates the loss of its rate-limiting window class at lower cost.

\section{Graph-Laplacian eigenvalue spectrum}\label{si:laplacian}
The migration network is an undirected weighted graph. With nonnegative edge weights $w_{uv}$
(taken as a decreasing function of the edge barrier, so that low-barrier hops are strongly
connected), the weighted degree is $D_{uu}=\sum_{v}w_{uv}$ and the combinatorial Laplacian is
\begin{equation}
L = D - W, \qquad W_{uv}=w_{uv},
\label{eq:laplacian}
\end{equation}
a symmetric positive-semidefinite operator whose eigenvalues
$0=\lambda_1\le\lambda_2\le\cdots\le\lambda_n$ encode global connectivity. The number of zero
eigenvalues equals the number of connected components, so the first nonzero eigenvalue, the
algebraic connectivity $\lambda_2$, is strictly positive only when the network is connected
and grows with the density of independent connecting paths. The associated Fiedler
eigenvector partitions the network at its weakest interface, and the Cheeger inequality bounds
the sparsest cut from above and below by $\lambda_2$, so $\lambda_2$ is a quantitative,
basis-free measure of how cross-linked the conduction network is.

Two descriptors are extracted from this spectrum. The \emph{network density} is $\lambda_2$ of
the resting-site (mobile-only) sublattice; because it is an eigenvalue of an abstract weighted
graph, it is invariant under a shift of the cell origin and quantifies the redundancy of the
mobile network independently of cell-size conventions. The \emph{saddle--cut consistency}
compares the location of the minimax saddle, Eq.~\eqref{eq:minimax}, with the Fiedler minimum
cut of $\hat G$: coincidence indicates a single rate-limiting interface (a bridge), whereas
separation indicates that bypasses exist (a cycle). The eigenvalue $\lambda_2$ and the
connectivity-transition shape are reported as origin-invariant quantities; the Fiedler
eigenvector itself is non-unique when $\lambda_2$ is degenerate, so cut-\emph{position} claims
are made only when the $\lambda_3-\lambda_2$ gap is nonzero. Eigenpairs are obtained with a
sparse symmetric solver in shift-invert mode for numerical stability.

\section{Persistent homology}\label{si:ph}
Persistent homology records how topological features of the network appear and merge as a
threshold is raised, providing a multiscale, basis-free count of connectivity and loops. The
filtration is built on the void-augmented network by admitting edges in order of increasing
barrier rather than by void radius or distance: at barrier $\beta$ the complex $K_\beta$
contains every edge with $B_{uv}\le\beta$, and $K_\beta\subseteq K_{\beta'}$ for
$\beta\le\beta'$. The homology groups of $K_\beta$ are tracked across the filtration. The
zeroth group $H_0$ counts connected components, whose mergers as $\beta$ rises trace the
assembly of the percolating network. The first group $H_1$ counts independent loops; each loop is born at the threshold that closes it, and the collection of births forms the loop-birth spectrum.

Because the network is periodic, a distinction is drawn between contractible loops, which
close within a single cell, and \emph{winding} loops, which close only through a periodic
image and therefore correspond to percolating cycles that span the crystal. The first
winding-loop birth coincides with the minimax percolation barrier $B_{\mathrm{perc}}$ by
construction, since both orderings are induced by the same edge barrier $B_{uv}$. This
coincidence is used as an internal consistency check on the bottleneck and the transport
dimensionality, and the signed difference between the two barriers is not reported as a
descriptor: it is identically zero for a correctly constructed filtration, and the physical
question it would otherwise address, whether the rate-limiting hop is bypassable, is answered
by the $B^\ast$-to-saddle ordering, the ablation sequence of Section~\ref{si:redundancy}, and $\lambda_2$. The transporting-loop (winding) fraction and the loop-birth spectrum are the homological outputs retained; the total number of features and
the winding subset are stable in count, whereas a winding/total ratio is sensitive to near-degenerate barriers and is not reported as a quantitative value. The present filtration
uses minimum-image distances, which suffices for the ordinal, within-structure claims made here; a fully rigorous periodic persistent homology is left as a refinement.


\section{The Bloch electronic spectrum and the dual network}\label{si:bloch}
A parallel electronic network is constructed to classify whether a material can carry
electrons alongside ions. Its edges are transition-metal--anion--transition-metal
superexchange links, whose strength follows Goodenough--Kanamori geometry: the
metal--anion--metal bond angle and the metal--anion distances set the overlap that governs
electron hopping, so a $180^\circ$ corner-sharing geometry allows stronger transfer than a
$90^\circ$ edge-sharing one. This network is weighted accordingly and, like the ionic network,
is analyzed through its Laplacian. Because the underlying lattice is periodic, the electronic
connectivity is evaluated as a Bloch (periodic) Laplacian, and the algebraic connectivity
$\lambda_2^{\mathrm{elec}}$ of the transition-metal sublattice measures whether a spanning
electronic pathway exists.

The dual classification follows from the two networks. A material whose ionic network
percolates while no transition-metal electronic network is present is an \emph{ideal
electrolyte}; one in which both the ionic and electronic networks percolate is a \emph{mixed
conductor}; one in which the ionic network is walled is an \emph{insulator}. For a mixed
conductor, the spatial separation between the ionic minimum cut and the electronic minimum cut
indicates whether the two limitations are coupled or independent: a large separation marks
decoupled limitations addressable by separate handles, as for \ce{LiFePO4}, where the ionic
and electronic cuts are separated by several \aa ngstr\"om.

\section{Vacancy response}\label{si:vacancy}
A dilute model introduces a single vacancy and records, per axis, the clearance and whether a
spanning route appears. A snapshot model populates the mobile sublattice at a vacancy fraction
$x$ on a grid, drawing several random configurations per fraction and averaging the per-axis
open probability and the mean number of open axes; the random sampling is seeded so that a run
is reproducible, and the onset fraction and the maximum dimensionality unlocked are insensitive to
the seed while the saturation detail carries minor sampling noise. The structure is evaluated
at its real composition $x_{\mathrm{real}}$. A structure that percolates only for $x>0$ is
vacancy-mandatory; the two cathodes here require delithiation to percolate, and the ordered
garnet opens nothing at its fully occupied composition. A per-class variant vacates one
symmetry-inequivalent mobile-site class at a time, the others held fully occupied, and ranks
the classes by the dimensionality each opens and the fraction at which it does so; classes are
grouped by a frame-invariant local-environment fingerprint, so the partition is
origin-independent.

\section{Route serialization and frame annotation}\label{si:frame}
The predicted route is serialized as an ordered list of nodes with periodic-image unwrapping,
together with the index of the rate-limiting edge. The coordinates are in the analysis frame,
the frame in which the engine built the graph, which may differ from an input CIF origin
setting. Each node is annotated, in the analysis frame, with its nearest crystallographic atom
(species and distance), and the frame is recorded explicitly. This makes each node
self-describing: a route node resolves to its mobile ion at $\approx 0$~\AA, a void resolves to
the framework atoms it sits among, and no node sits within bonding distance of a framework
cation. This node-level check is a required validation gate; comparing serialized coordinates
against a structure in a different origin setting is unsafe and is not done.

\section{Datasets and computational details}\label{si:compute}
Structures were taken from the Materials Project~\cite{Jain2013_si} and from the deposited
crystallographic data of the cited diffraction studies, and were validated against their
reported space groups before analysis. The validation references are the AIMD, NEB, and
diffraction studies cited in the main text for each structure. Structure handling used
pymatgen.~\cite{Ong2013_si} The interstitial network used a Voronoi seeding; spectral
descriptors used sparse symmetric eigensolvers in shift-invert mode; persistent homology was
computed on the energy-filtered network. Each structure is analyzed in seconds on a single
processor, with no electronic-structure or molecular-dynamics calculation in the loop. All
descriptor values in Tables~\ref{tab:descriptors}--\ref{tab:vacancy} were extracted directly from
the network-topology and descriptor outputs of the current code version, and serialized routes
carry frame-explicit, self-describing node annotations.

\paragraph{Symmetry validation.} A migration barrier is a function of the crystal, so
hops related by a space-group operation are required to carry identical barriers. This
condition is verified rather than assumed. For each structure the rate-limiting edge is
mapped through every symmetry operation of the parent group, both endpoints are matched
back to network nodes, and the barriers of the resulting image edges are compared. The
number of image edges recovered equals the group order in every case ($16$ for
\ce{Li10GeP2S12}, $288$ for \ce{LiCoO2}, $96$ for \ce{Li6PS5Cl}), and the barrier
spread across the orbit vanishes for all eight structures to within $10^{-3}$~meV,
which is the double-precision accumulation limit for a multi-electronvolt sum over
$10^{4}$ framework atoms. The check is inexpensive and precedes the reporting of any
descriptor. A non-vanishing spread indicates that the constructed network does not
inherit the symmetry of the parent cell, and invalidates every quantity derived from
it, including the percolation barrier, the ablation increment, and the degeneracy.

\section{Per-structure descriptor tables}\label{si:tables}
Tables~\ref{tab:descriptors}--\ref{tab:vacancy} list the descriptor values, serialized routes,
and vacancy response extracted directly from the network-topology and descriptor outputs of the
current code version.

\begin{table*}[t]
\centering
\caption{Transport-mechanism descriptors across the validation set, computed from a
single static structure. $B_{\mathrm{perc}}$, $B^\ast$, the anisotropy, and $\Delta B_1$
are ordinal: they order directions and hops \emph{within} a structure and are not
comparable in magnitude across rows. $B_{\mathrm{perc}}$, Eq.~\eqref{eq:minimax}, is the
minimax percolation barrier on the void-augmented network. $B^\ast$,
Eq.~\eqref{eq:bstar}, is the lowest-barrier direct mobile-to-mobile hop, evaluated on
the pristine resting-site graph without void decoration; the ordering of $B^\ast$
against $B_{\mathrm{perc}}$ separates direct channel transport (negative difference)
from interstitial-mediated transport (positive difference). The anisotropy is
$\max_a\Delta_a-\min_a\Delta_a$ over the axis-opening thresholds,
Eq.~\eqref{eq:winding}. Degeneracy is the symmetry multiplicity of the rate-limiting
window, the number of crystallographically identical copies carrying identical
barriers, given for the analysed cell with the value per primitive cell in parentheses;
it confers point-defect tolerance only in combination with a conduction dimensionality
of at least two. $\Delta B_1$ is the barrier increment to the next thermally distinct
percolating route once the rate-limiting orbit is removed, measured against the global
$B_{\mathrm{perc}}$. $\lambda_2^{\mathrm{mob}}$ is the unweighted algebraic
connectivity of the mobile sublattice after the steric-wall tail is removed; it is
extensive and is reported with $N_{\lambda_2}$, the number of nodes in the component on
which it is evaluated. The barrier spread across the symmetry orbit of the
rate-limiting edge vanishes for every row (Section~\ref{si:compute}).
$^{\ast}$Geometrically three-dimensional but occupancy-limited at full occupancy.
$^{\dagger}$$\lambda_2$ evaluated on a two-node component: the trivial eigenvalue of a
single edge, carrying no information on network density.}
\label{tab:descriptors}
\small
\setlength{\tabcolsep}{4pt}
\begin{tabular}{lccccccrrr}
\toprule
 & & \multicolumn{4}{c}{Barriers (eV)} & Degeneracy & Redund. & \multicolumn{2}{c}{Network density} \\
\cmidrule(lr){3-6}\cmidrule(lr){7-7}\cmidrule(lr){8-8}\cmidrule(lr){9-10}
Structure & Dim. & $B_{\mathrm{perc}}$ & $B^\ast$ & $B^\ast{-}B_{\mathrm{perc}}$ & Aniso. & (per prim.) & $\Delta B_1$ (meV) & $\lambda_2^{\mathrm{mob}}$ & $N_{\lambda_2}$ \\
\midrule
\ce{LiFePO4}           & 1 & 0.185 & 0.035 & $-0.150$ & 0.000 & 64 (8) & 296 & 2.00$^{\dagger}$ & 2 \\
\ce{LiCoO2}            & 2 & 7.191 & 10.674 & $+3.483$ & 0.000 & 144 (6) & 3483 & 4.59 & 48 \\
\ce{Li10GeP2S12}       & 3 & 3.394 & 0.567 & $-2.827$ & 0.000 & 16 (16) & 83 & 2.00$^{\dagger}$ & 2 \\
\ce{Na3PS4} (cubic)    & 3 & 2.318 & 2.318 & $+0.000$ & 0.000 & 12 (12) & 1624 & 1.17 & 12 \\
\ce{Na3PS4} (tet.)     & 3 & 2.806 & 1.090 & $-1.716$ & 2.339 & 8 (8) & 275 & 1.37 & 12 \\
\ce{Li6PS5Cl}          & 3 & 0.102 & 7.811 & $+7.708$ & 0.000 & 96 (24) & 710 & 6.00 & 24 \\
tet-\ce{Li7La3Zr2O12}  & 3$^{\ast}$ & 4.788 & 5.758 & $+0.970$ & 0.359 & 32 (16) & 359 & 9.84 & 64 \\
$\gamma$-\ce{LiAlO2}   & 1 & 1.801 & 0.150 & $-1.651$ & 0.000 & 64 (8) & 1579 & 2.00$^{\dagger}$ & 2 \\
\bottomrule
\end{tabular}
\\[3pt]
\footnotesize Analysed cell $N_{\mathrm{cell}}$ (sites, primitive multiples):
\ce{LiFePO4} 224 ($8\times$), \ce{LiCoO2} 96 ($24\times$), \ce{Li10GeP2S12} 70
($1\times$), \ce{Na3PS4} cubic 16 ($2\times$), \ce{Na3PS4} tet.\ 16 ($1\times$),
\ce{Li6PS5Cl} 52 ($4\times$), tet-\ce{Li7La3Zr2O12} 192 ($2\times$),
$\gamma$-\ce{LiAlO2} 128 ($8\times$).
\end{table*}

\begin{table*}[t]
\centering
\caption{Decomposition of the rate-limiting saddle into the three parameter-free
contributions of Eq.~\eqref{eq:edgebarrier}, for the lowest-barrier open axis of each
structure. The values are the contributions evaluated at the saddle, before subtraction
of the site baseline $E_0$; $B_{\mathrm{perc}}$ is given for reference and is not the
sum of the three columns. The final column is the ratio of the steric to the
electrostatic contribution and measures which physics selects the route. The steric
Born--Mayer contribution exceeds the electrostatic contribution by between one and two
orders of magnitude throughout, so the migration barrier is governed by the aperture
presented by the framework rather than by the charge distributed across it. The
induced-dipole contribution is uniformly stabilizing. $\gamma$-\ce{LiAlO2} carries the
only positive electrostatic contribution in the set, and is the only structure for
which electrostatics opposes migration.}
\label{tab:decomp}
\begin{tabular}{lcrrrrr}
\toprule
Structure & Axis & $E_{\mathrm{ster}}$ & $E_{\mathrm{elec}}$ & $E_{\mathrm{ind}}$ & $B_{\mathrm{perc}}$ & $E_{\mathrm{ster}}/|E_{\mathrm{elec}}|$ \\
 & & (eV) & (eV) & (eV) & (eV) & \\
\midrule
\ce{LiFePO4}           & $b$ & 26.985 & $-1.434$ & $-4.584$ & 0.185 & 19 \\
\ce{LiCoO2}            & $a$ & 25.550 & $-0.156$ & $-5.897$ & 7.191 & 164 \\
\ce{Li10GeP2S12}       & $b$ & 11.767 & $-0.133$ & $-5.032$ & 3.394 & 88 \\
\ce{Na3PS4} (cubic)    & $c$ & 10.825 & $-0.118$ & $-4.128$ & 2.318 & 92 \\
\ce{Na3PS4} (tet.)     & $a$ & 10.205 & $-0.077$ & $-4.016$ & 2.806 & 132 \\
\ce{Li6PS5Cl}          & $b$ &  9.333 & $+0.018$ & $-4.128$ & 0.102 & 507 \\
tet-\ce{Li7La3Zr2O12}  & $c$ & 18.816 & $-3.990$ & $-4.263$ & 4.788 & 5 \\
$\gamma$-\ce{LiAlO2}   & $c$ & 17.248 & $+3.108$ & $-3.943$ & 1.801 & 6 \\
\bottomrule
\end{tabular}
\end{table*}

\begin{table}[htbp]
\centering
\caption{Serialized percolation routes with analysis-frame framework shells, extracted from the
analysis outputs. ``Open axes'' are the axes that acquire a spanning route on the
resting-site network. The framework shell lists the closest framework contacts that border
the rate-limiting saddle. The \ce{Li10GeP2S12} \ce{Li1}/\ce{Li3} sub-labels are assigned by
Wyckoff occupancy.}
\label{tab:routes}
\small
\begin{tabular}{llll}
\toprule
Structure & Open axes & Route (bottleneck $\star$) & Framework shell at saddle (\AA) \\
\midrule
\ce{LiFePO4}       & $b$     & Li$\star$Li along [010] (curved) & O 1.74, O 2.41, Fe 3.13 \\
\ce{LiCoO2}        & $a,b$   & Li$\star$Li in-plane (curved) & O 1.73, O 1.73, O 3.05 \\
\ce{Li10GeP2S12}   & $a,b,c$ & \ce{Li1}-\ce{Li1}-\ce{Li1}$\star$\ce{Li3}-\ce{Li3}-\ce{Li1}-\ce{Li1} & S 2.32, S 2.33, Ge 3.29 \\
\ce{Na3PS4} (cubic)& $a,b,c$ & saddle at $12d\ (\tfrac14,0,\tfrac12)$, $6b\!\to\!12d\!\to\!6b$ & S 2.63, S 2.63, P 3.88 \\
\ce{Na3PS4} (tet.) & $a,b,c$ & Na$\star$Na (curved $a,b$; $c$ hardest) & S 2.57, S 2.57, S 2.79 \\
\ce{Li6PS5Cl}      & $a,b,c$ & $24g$(\ce{T5a})$\star$void$\to$$24g$ inter-cage window & S 2.25, S 2.36, S 2.68, Cl 3.30 \\
tet-\ce{Li7La3Zr2O12} & $a,b,c^{\ast}$ & $8a$(tet)$\to$$32g$(oct)$\star$ O-bordered (occupancy-walled) & O 1.76, O 1.83, O 1.83 \\
$\gamma$-\ce{LiAlO2}  & $c$  & Li$\star$Li along [001] (curved) & O 1.71, O 1.78, O 2.36 \\
\bottomrule
\end{tabular}

\vspace{2pt}
\footnotesize $^{\ast}$Geometric connectivity only; no single-ion route at full occupancy.
\end{table}

\begin{table}[htbp]
\centering
\caption{Vacancy response. ``Stoich.\ percolates'' is percolation at the ideal composition;
``Dilute axes'' lists the axes opened once vacancies are present; ``Percolates at
$x_{\mathrm{real}}$'' is the verdict at the real operating composition. The cathodes are
evaluated from the fully lithiated endpoint, so percolation requires delithiation.}
\label{tab:vacancy}
\small
\begin{tabular}{lccc}
\toprule
Structure & Stoich.\ percolates & Dilute axes opened & Percolates at $x_{\mathrm{real}}$ \\
\midrule
\ce{LiFePO4}          & no  & $b$       & no (needs delithiation) \\
\ce{LiCoO2}           & no  & $a,b$     & no (needs delithiation) \\
\ce{Li10GeP2S12}      & no  & $a,b,c$   & yes (operating composition) \\
\ce{Na3PS4} (cubic)   & no  & $a,b,c$   & no (vacancy-mandatory) \\
\ce{Na3PS4} (tet.)    & no  & $a,b,c$   & no (vacancy-mandatory) \\
\ce{Li6PS5Cl}         & no  & $a,b,c$   & no (disorder/vacancy-mandatory) \\
tet-\ce{Li7La3Zr2O12} & no  & none      & no (opens nothing) \\
$\gamma$-\ce{LiAlO2}  & no  & $c$ only  & no ($c$ opens under vacancy) \\
\bottomrule
\end{tabular}
\end{table}

\noindent\textbf{Cathode dual-network detail.} \ce{LiFePO4}: ionic network limiting, the
electronic network carried by \ce{Fe}; the ionic and electronic minimum cuts are spatially
separated (decoupled limitations). \ce{LiCoO2}: electronic network carried by \ce{Co}, ionic
network limiting. The sulfide electrolytes carry no transition-metal electronic network (ideal
electrolytes), and the ordered oxides are insulators at their real composition.

\section{Route validation status}\label{si:validation}
Each serialized route was checked against the deposited structure with spglib, assigning every
resting node to a Wyckoff orbit, and against the mechanism reported by independent AIMD, NEB,
or diffraction (Table~\ref{tab:validation}). Special-position saddles ($12d$, $6b$) are
confirmable from fractional coordinates alone; parametric orbits ($16h$, $24g$, $8a$/$32g$)
were assigned by spglib on the supplied CIF. The cubic \ce{Na3PS4} $6b\!\to\!12d\!\to\!6b$
route is confirmed on all three axes. The \ce{Li10GeP2S12} route is confirmed as two
interleaved $16h$ orbits (the \ce{Li1} and \ce{Li3} channel sites) cross-linked through a $4c$
(\ce{Li4}) site, with the rate-limiting \ce{Li1}$\to$\ce{Li3} hop bordered by sulfur. The
argyrodite resting sites resolve onto the $24g$ (\ce{T5a}) orbit of an ordered model; its full
\ce{T5}/\ce{T5a}/\ce{T2} inter-cage sequence and the order-to-disorder dependence are not
represented by a single ordered structure. The $\gamma$-\ce{LiAlO2} route runs along the Li
sites of the $c$-channel in the analysis frame, but that frame differs from the deposited
enantiomorphic cell, so its node-level Wyckoff assignment is pending a frame reconciliation;
the dimensionality and channel character are secure.

\begin{table}[htbp]
\centering
\caption{Route validation status. ``Method'' is the means of node assignment; ``Status'' is
the agreement with the reference mechanism. For \ce{LiCoO2}, TSPD reproduces the tetrahedral
activated state; the divacancy character and the concentration dependence of the barrier are
dynamical and configurational effects on the rate, outside a static treatment.}
\label{tab:validation}
\small
\begin{tabular}{lll}
\toprule
Structure & Verified route / saddle & Status \\
\midrule
\ce{LiFePO4}          & curved [010] channel, O-bordered & matches neutron \ce{Li}$_{0.6}$\ce{FePO4} \\
\ce{LiCoO2}           & in-plane, tetrahedral activated state & matches activated state (see caption) \\
\ce{Na3PS4} (cubic)   & $6b\!\to\!12d\!\to\!6b$ (all axes, spglib) & confirmed \\
\ce{Na3PS4} (tet.)    & $4d/2a$-derived resting, low-symmetry saddle & confirmed (symmetry split) \\
\ce{Li10GeP2S12}      & $16h$(\ce{Li1})/$16h$(\ce{Li3}) + $4c$ link (spglib) & confirmed; \ce{Li1}$\to$\ce{Li3} bottleneck \\
\ce{Li6PS5Cl}         & $24g$(\ce{T5a}) inter-cage window, ordered model (spglib) & mechanism confirmed; see scope \\
tet-\ce{Li7La3Zr2O12} & $8a\!\to\!32g$ O-bordered (spglib) & confirmed; occupancy-walled \\
$\gamma$-\ce{LiAlO2}  & $c$-channel along Li (spglib) & 1D confirmed; orbit labels pending frame \\
\bottomrule
\end{tabular}
\end{table}


\section{Full per-structure figure set}\label{si:figs}
This section collects, for each structure, the route decomposition, the migration
network, the vacancy response, and the dual ionic/electronic overlay.
Figure file stems follow the uploaded \texttt{Figures/} tree.

\subsection{Route decompositions (unified / geometric / electrostatic)}
\begin{figure}[htbp]
\centering
\includegraphics[width=1.0\linewidth]{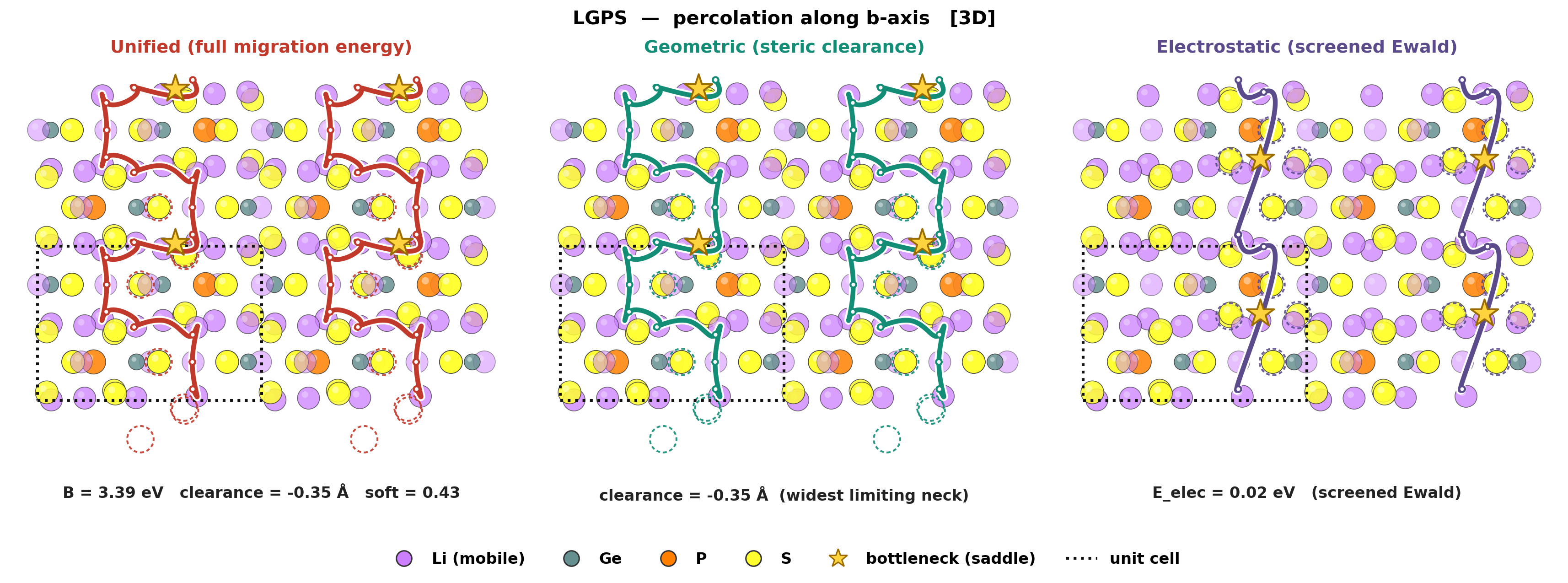}
\caption{Route decomposition for \ce{Li10GeP2S12}. Unified,
geometric-only, and electrostatic-only routes on a shared 2D projection, with the
framework-bordered bottleneck starred. Coincidence of the unified and geometric routes
identifies a steric wall; divergence of the electrostatic route shows that electrostatics do
not set the path.}
\label{fig:decomp1}
\end{figure}

\begin{figure}[htbp]
\centering
\includegraphics[width=1.0\linewidth]{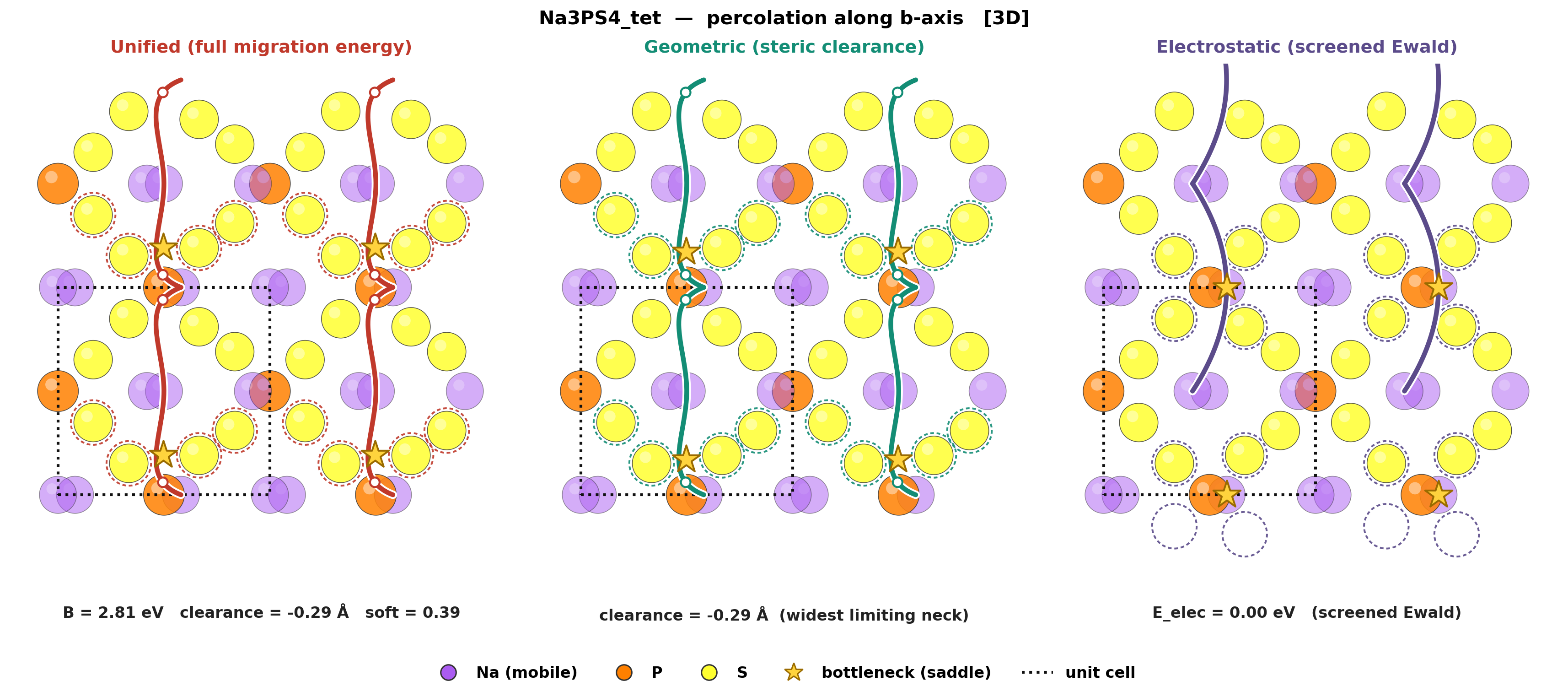}\\[6pt] 
\includegraphics[width=1.0\linewidth]{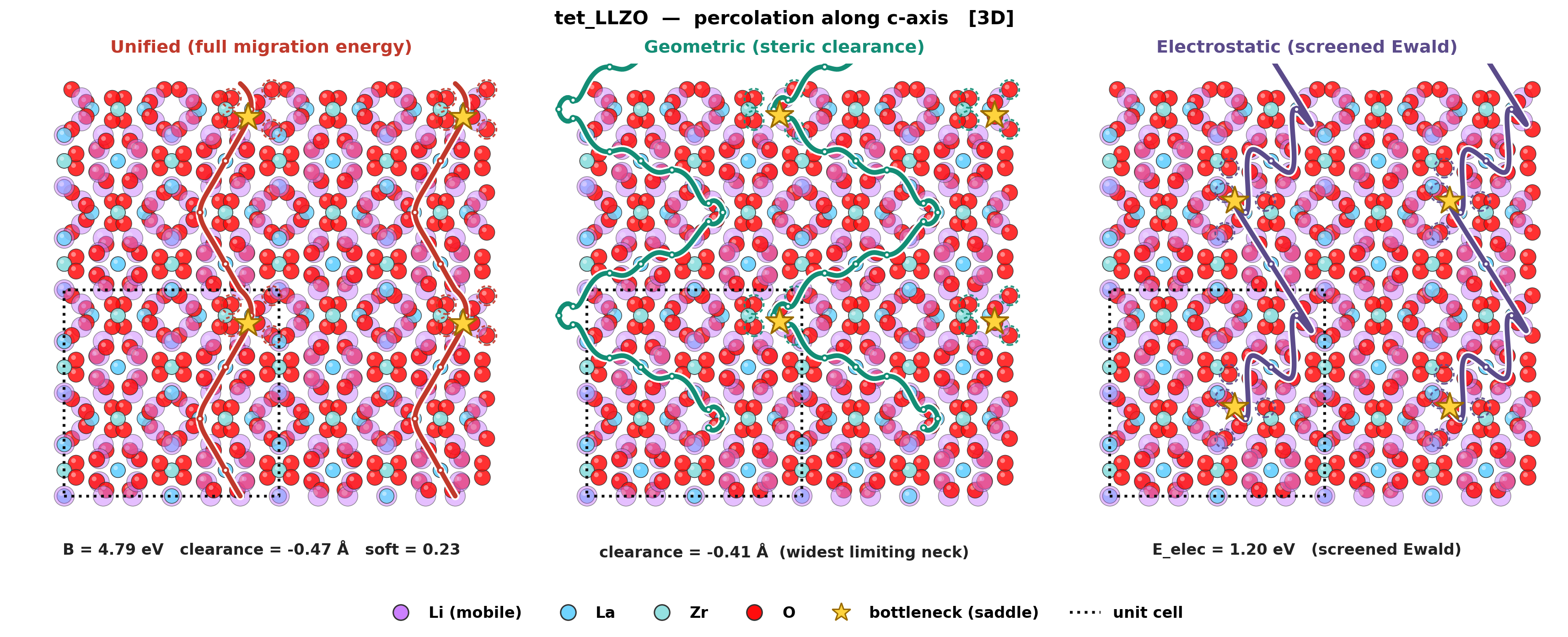}
\caption{Route decomposition for tetragonal \ce{Na3PS4} (top) and tetragonal
\ce{Li7La3Zr2O12} (bottom). For tetragonal \ce{Na3PS4} the $c$ axis bears the hardest window,
the structural fingerprint of the non-superionic polymorph; for the garnet the geometric
network is connected but the structure is occupancy-walled.}
\label{fig:decomp2}
\end{figure}

\subsection{Visualization of percolation paths}
\begin{figure}[htbp]
\centering
\begin{minipage}{0.48\linewidth}
    \leftline{\textbf{(a)}}
    \vspace{1pt}
    \includegraphics[width=1.0\linewidth]{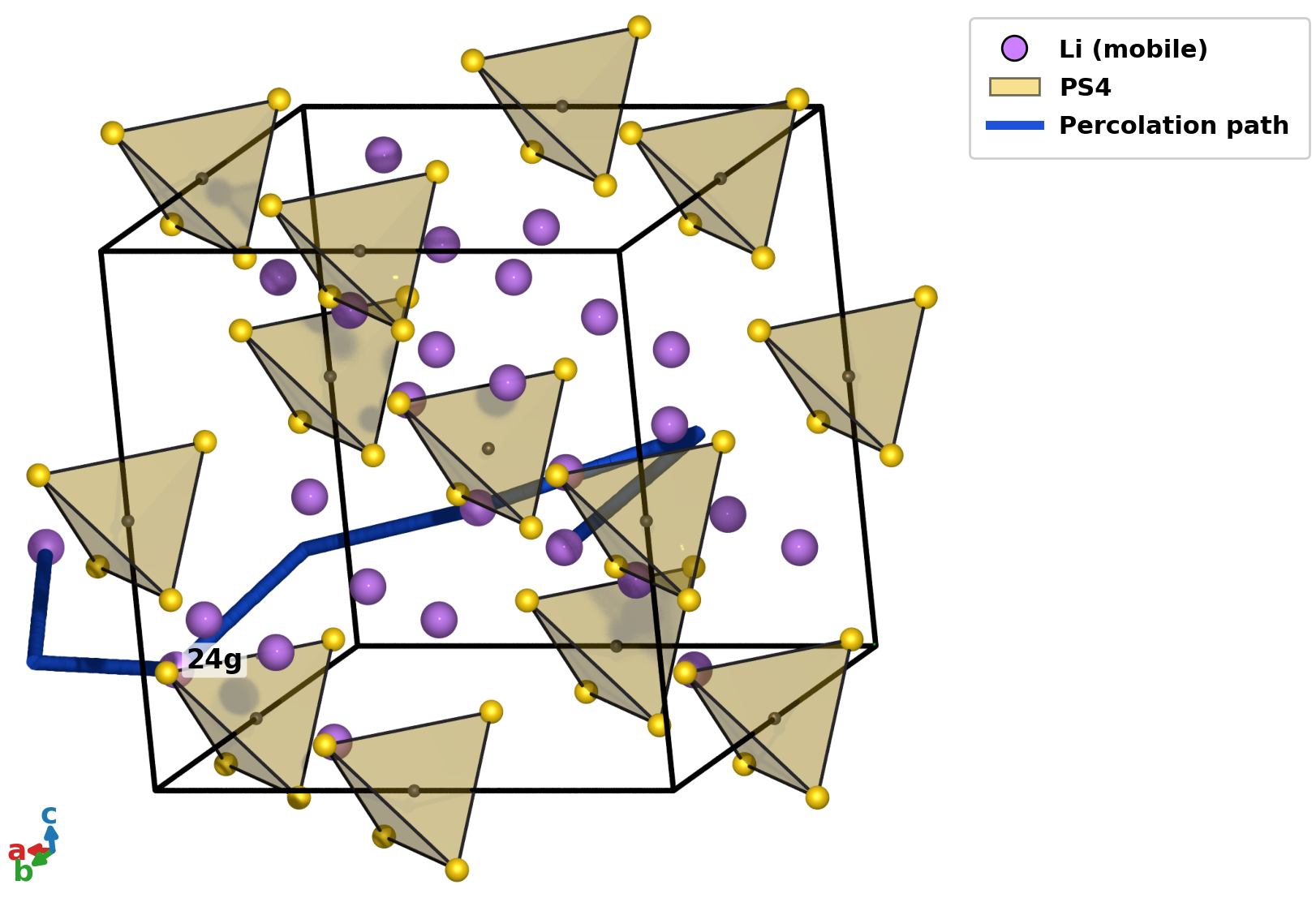}
\end{minipage}\hfill
\begin{minipage}{0.48\linewidth}
    \leftline{\textbf{(b)}}
    \vspace{1pt}
    \includegraphics[width=1.0\linewidth]{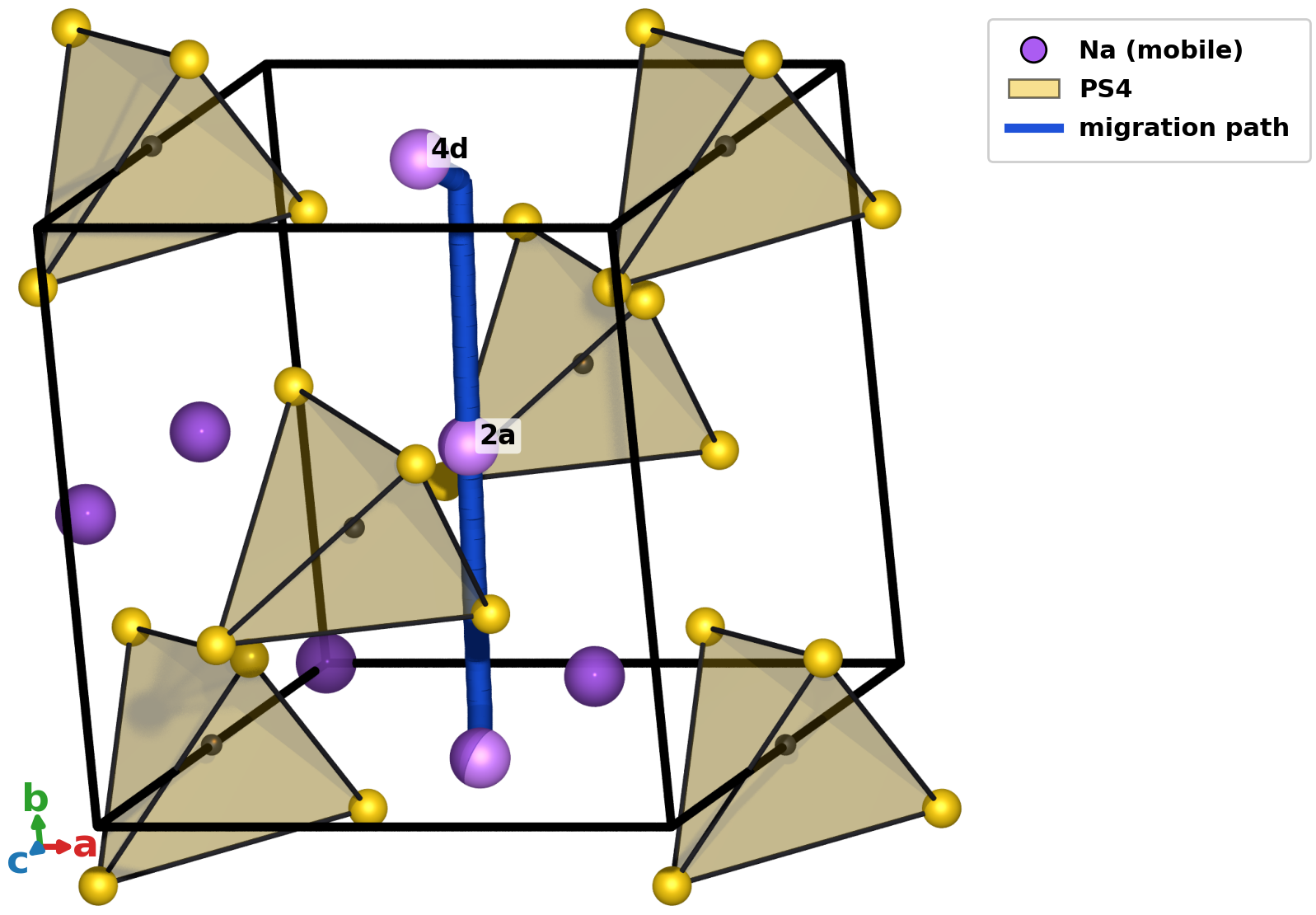}
\end{minipage}

\vspace{0.4cm} 

\begin{minipage}{0.48\linewidth}
    \leftline{\textbf{(c)}}
    \vspace{1pt}
    \includegraphics[width=1.0\linewidth]{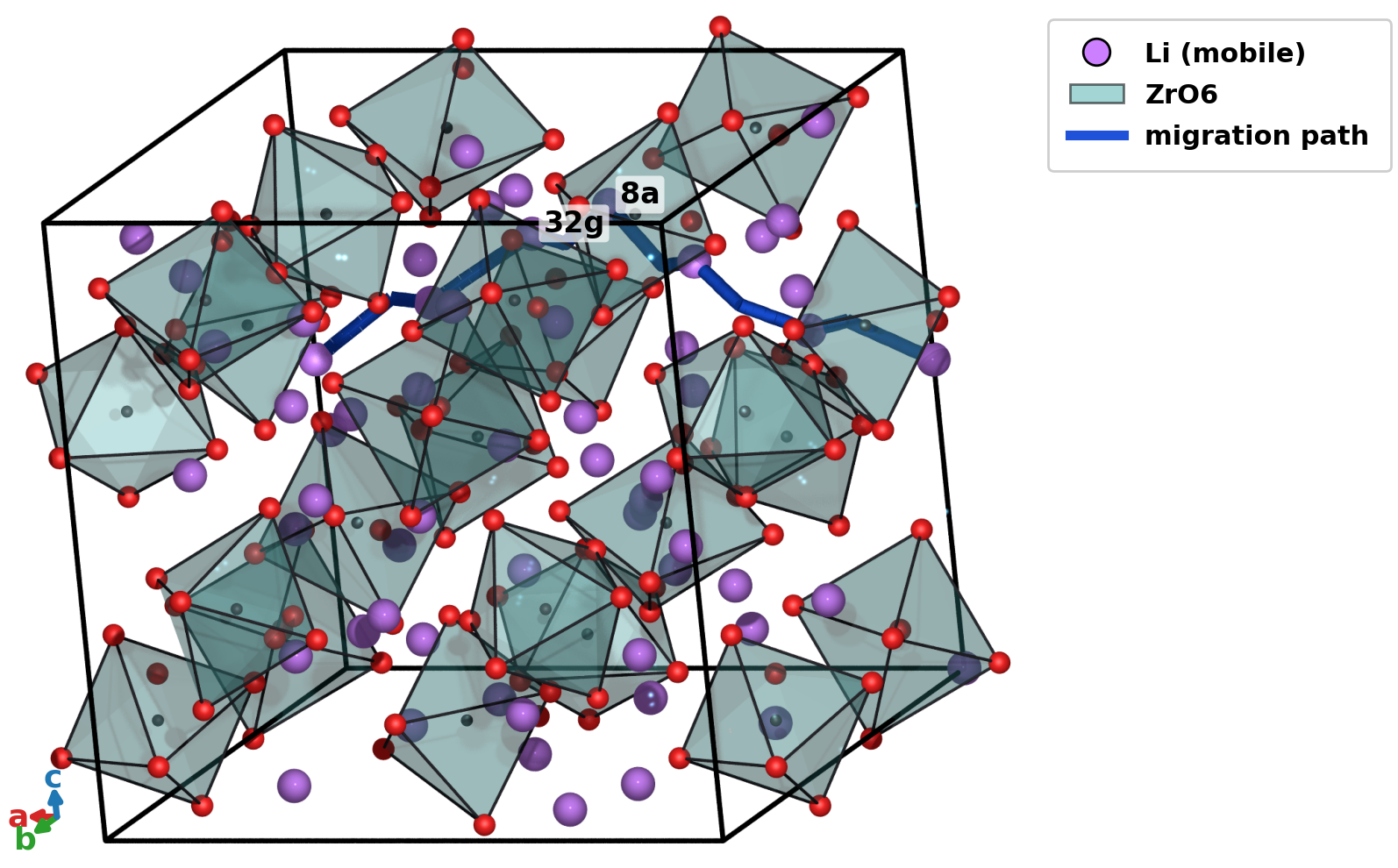}
\end{minipage}\hfill
\begin{minipage}{0.48\linewidth}
    \leftline{\textbf{(d)}}
    \vspace{1pt}
    \includegraphics[width=1.0\linewidth]{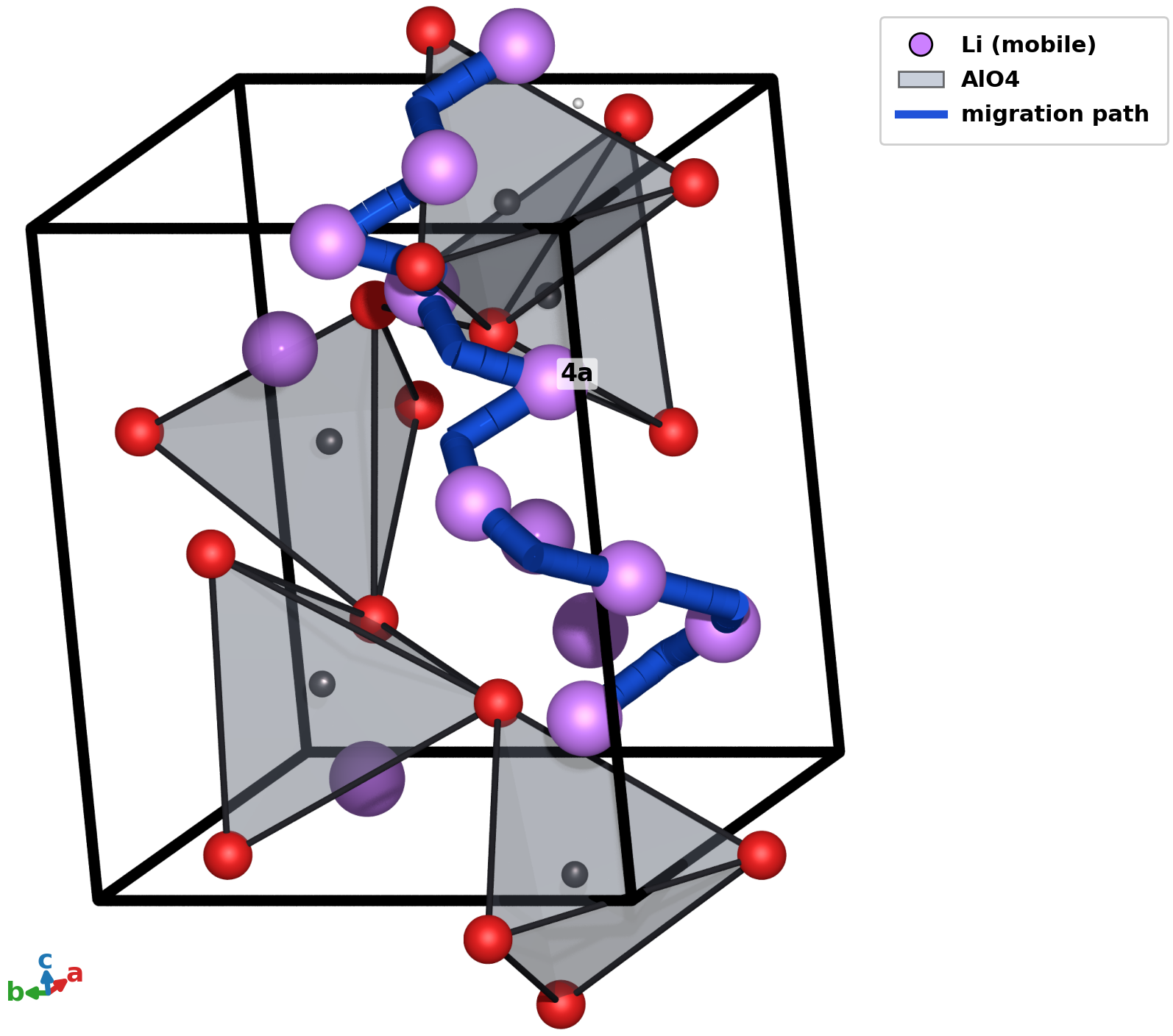}
\end{minipage}

\vspace{0.3cm}
\caption{Migration networks for \textbf{(a)} \ce{Li6PS5Cl}, \textbf{(b)} tetragonal \ce{Na3PS4}, \textbf{(c)} tetragonal \ce{Li7La3Zr2O12}, and \textbf{(d)} $\gamma$-\ce{LiAlO2}. The argyrodite is densely cross-linked; the garnet network is connected but occupancy-walled; $\gamma$-\ce{LiAlO2} shows the single $c$-channel.}
\label{fig:net3d2}
\end{figure}

\subsection{Vacancy response}
\begin{figure}[htbp]
\nopagebreak 
\vspace{-0.3cm} 
\centering
    \leftline{\hspace{0.075\linewidth}\textbf{(a)}}
    \includegraphics[width=0.90\linewidth]{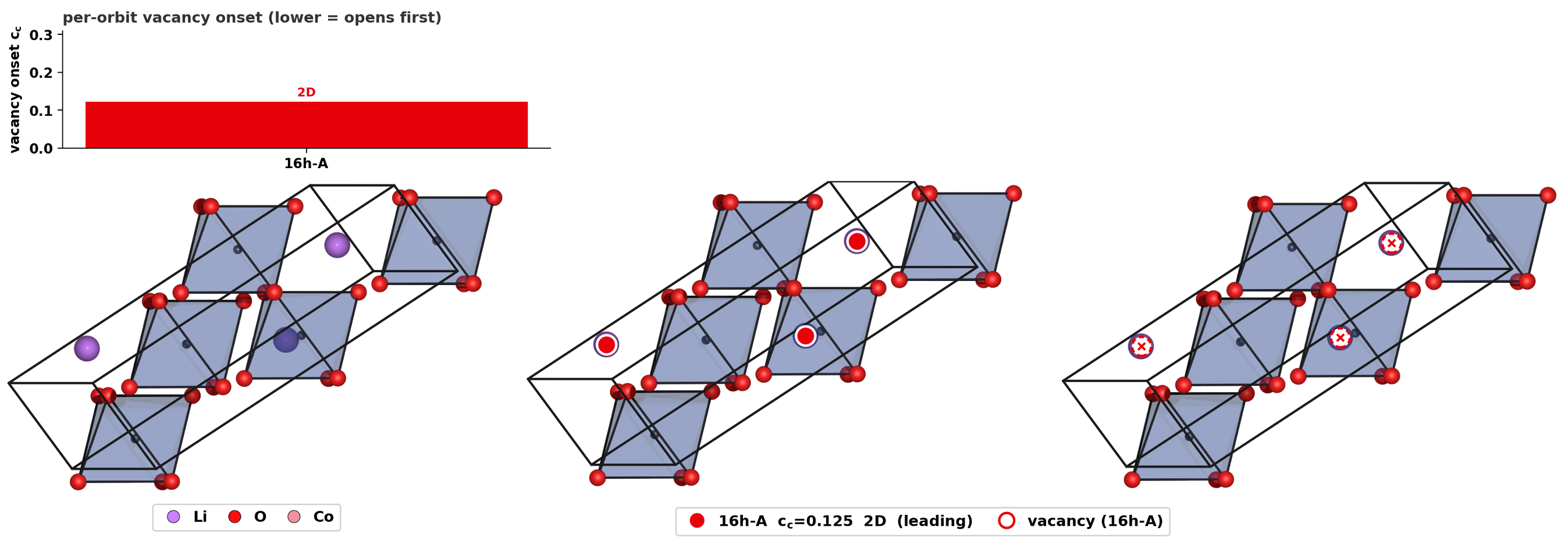}

    \vspace{0.15cm} 

    \leftline{\hspace{0.075\linewidth}\textbf{(b)}}
    \includegraphics[width=0.90\linewidth]{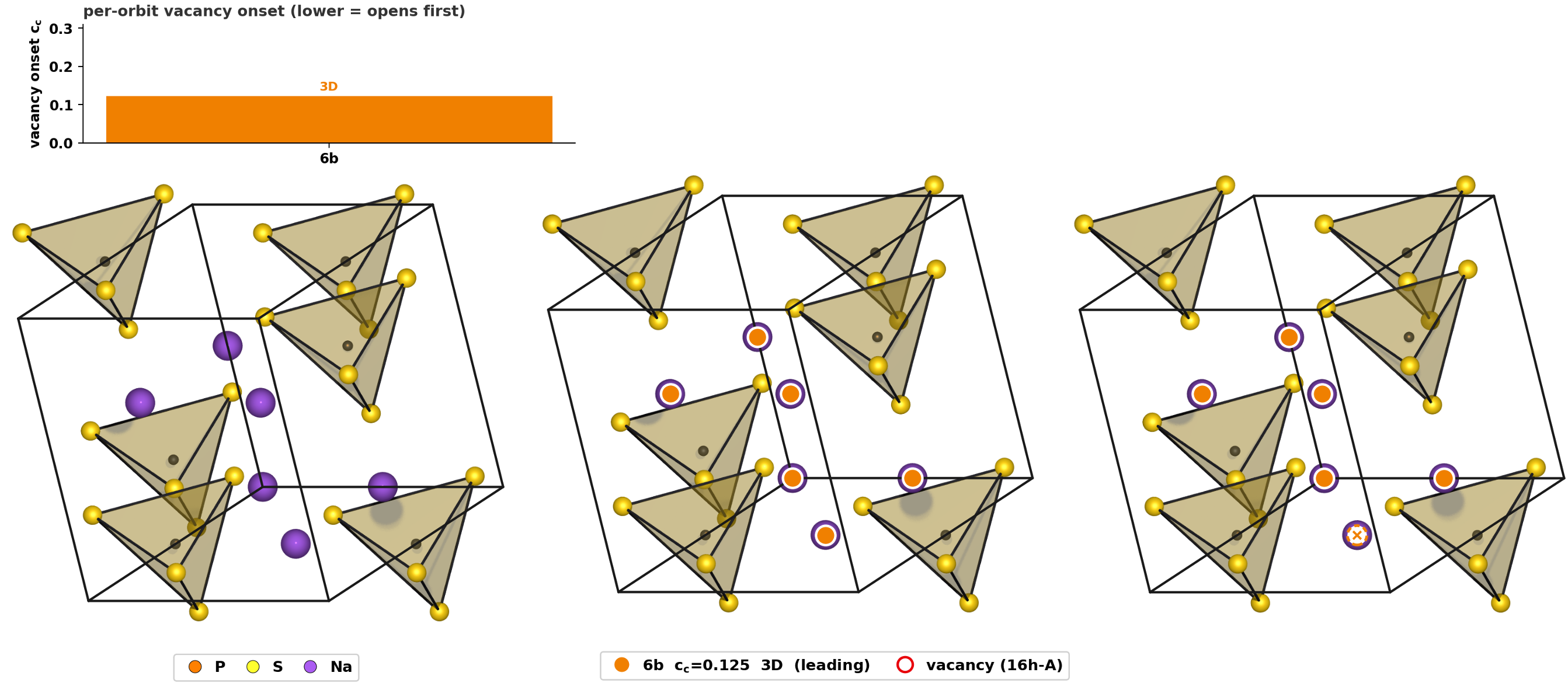}

    \vspace{0.15cm}

    \leftline{\hspace{0.075\linewidth}\textbf{(c)}}
    \includegraphics[width=0.90\linewidth]{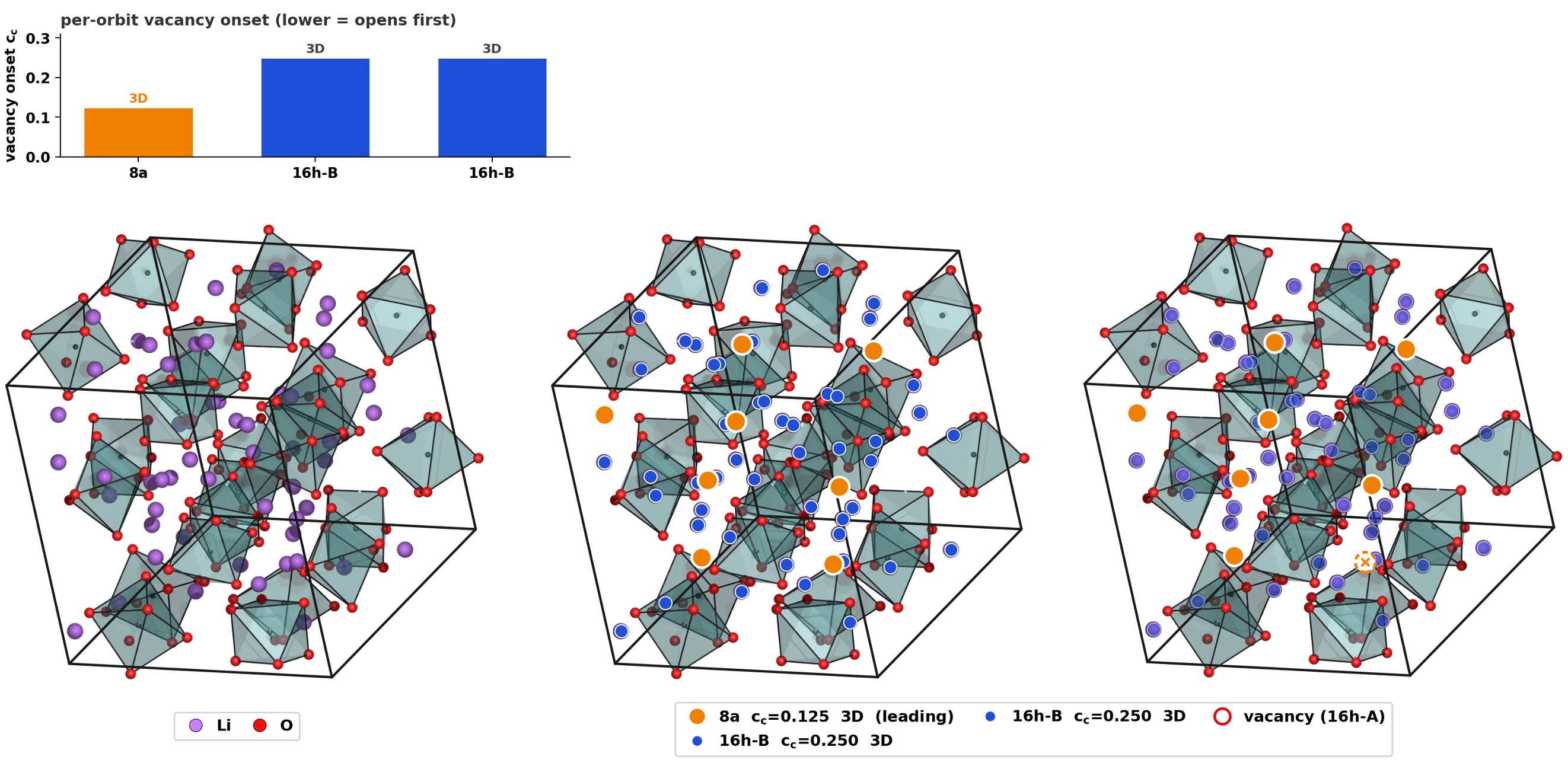}

\vspace{0.15cm}
\caption{Vacancy response for \textbf{(a)} \ce{LiCoO2} (delithiation opens the plane), 
\textbf{(b)} cubic \ce{Na3PS4} (vacancy-mandatory, all axes open with vacancies), and 
\textbf{(c)} tetragonal \ce{Li7La3Zr2O12} (opens nothing at the fully occupied composition). 
Per-axis percolation is plotted against vacancy fraction with the operating composition marked.}
\label{fig:vacancySI}
\vspace{-0.3cm} 
\end{figure}

\subsection{Dual ionic/electronic networks (cathodes)}
\begin{figure}[htbp]
\centering
\includegraphics[width=1.0\linewidth]{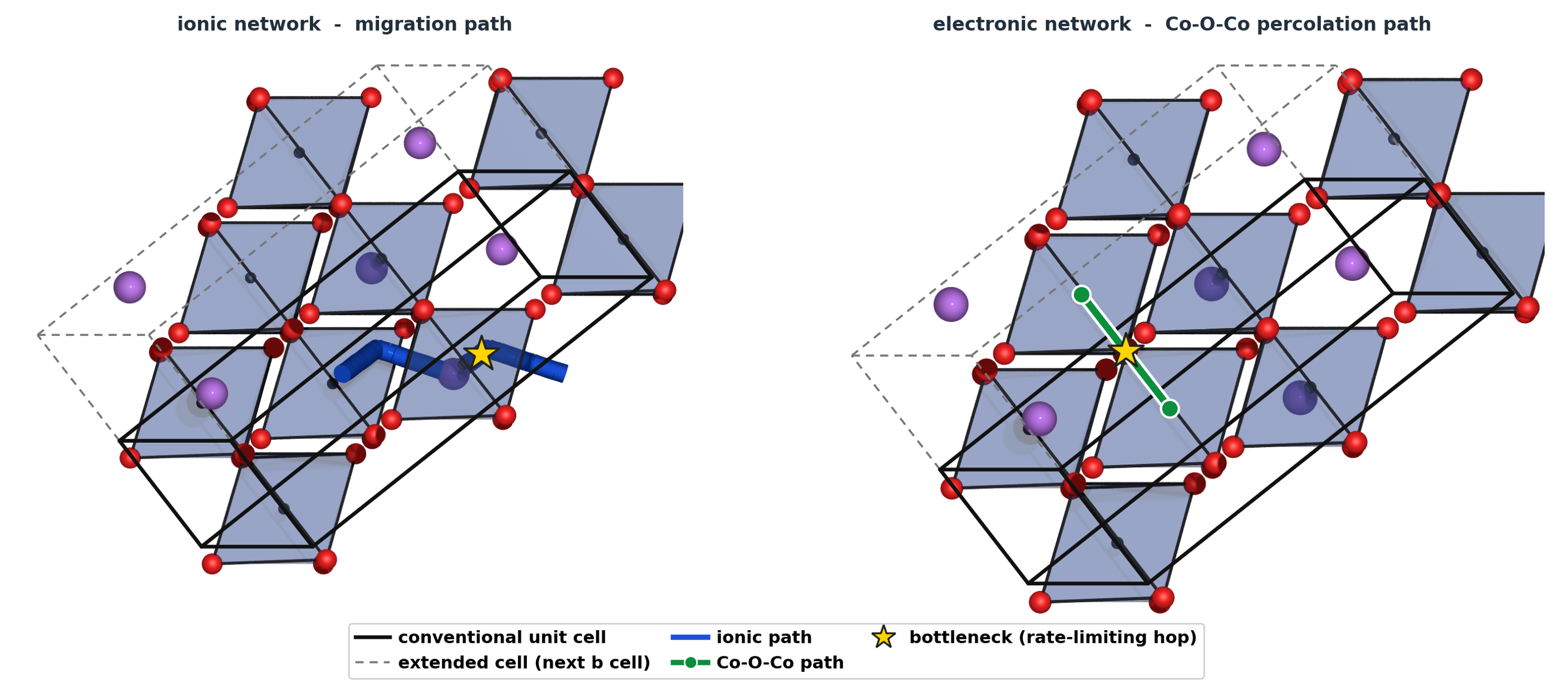}
\caption{Dual ionic/electronic bottleneck overlay for \ce{LiCoO2}. The transition-metal electronic network and the limiting ionic network are shown; for \ce{LiCoO2} the two minimum cuts are spatially separated, indicating decoupled design handles.}
\label{fig:dualSI}
\end{figure}